%% file: main.tex
\documentclass[prc, amsfonts, amssymb, amsmath, reprint, showkeys, nofootinbib]{revtex4-1}
\usepackage[english]{babel}
\usepackage[utf8]{inputenc}
\usepackage[colorinlistoftodos, color=green!40, prependcaption]{todonotes}
\usepackage{subfigure}
\input{preamble}
\usepackage{lineno}

\newcommand\pT{\ensuremath{p_\mathrm{T}}}
\newcommand\pTelectron{\ensuremath{p_\mathrm{T}^{e}}}

\newcommand\pTjet{\ensuremath{p_\mathrm{T}^{jet}}}

\newcommand\qhat{\ensuremath{\hat{q}}}
\newcommand\GeVsquare{\ensuremath{\mathrm{GeV}^{2}}}
\newcommand\Rg{\ensuremath{R_{g}}}

\newcommand\etalab{\ensuremath{\eta^{jet}}}
\newcommand\etacm{\ensuremath{\eta_{\mathrm{CM}}}}

\begin{document}
\title{Jets as precision probes in electron-nucleus collisions at the future Electron-Ion Collider}

\author{Miguel Arratia}
\email[Correspondence email address: ]{miguel.arratia@ucr.edu}
\altaffiliation{Current affiliation: Department of Physics and Astronomy, University of California, Riverside, CA 92521, USA and Thomas Jefferson National Accelerator Facility, Newport News, VA 23606, USA.}
\affiliation{Department of Physics, University of California, Berkeley, CA 94720, USA}
\affiliation{Nuclear Science Division, Lawrence Berkeley National Laboratory, Berkeley, CA 94720, USA}

\author{Youqi Song}
\affiliation{Department of Physics, University of California, Berkeley, CA 94720, USA}
\affiliation{Nuclear Science Division, Lawrence Berkeley National Laboratory, Berkeley, CA 94720, USA}

\author{Felix Ringer}
\affiliation{Department of Physics, University of California, Berkeley, CA 94720, USA}
\affiliation{Nuclear Science Division, Lawrence Berkeley National Laboratory, Berkeley, CA 94720, USA}

\author{Barbara V. Jacak}
\affiliation{Department of Physics, University of California, Berkeley, CA 94720, USA}
\affiliation{Nuclear Science Division, Lawrence Berkeley National Laboratory, Berkeley, CA 94720, USA}

\date{\today} 

\begin{abstract}
We discuss the prospects of using jets as precision probes in electron-nucleus collisions at the future Electron-Ion Collider. Jets produced in deep-inelastic scattering can be calibrated by a measurement of the scattered electron. Such electron-jet ``tag and probe'' measurements call for an approach that is orthogonal to most HERA jet measurements as well as previous studies of jets at the future EIC. We present observables such as the electron-jet momentum balance, azimuthal correlations and jet substructure, which can provide constraints on the parton transport coefficient in nuclei. We compare simulations and analytical calculations and provide estimates of the expected medium effects. Implications for  detector design at the future EIC are discussed. 
\end{abstract}

\maketitle

\section{Introduction} \label{sec:outline}

The future Electron-Ion Collider (EIC) will be the first electron-nucleus (e-A) collider and will produce the first jets in nuclear deep-inelastic scattering (DIS). Jet measurements can extend traditional semi-inclusive DIS to elucidate parton-nucleus interactions, the 3D structure of nuclei, and the parton-to-hadron transition, which are among the physics goals of the future EIC~\cite{Accardi:2012qut}. 

Most studies discussed in the future EIC white paper~\cite{Accardi:2012qut} are based on single-hadron measurements. But since 2011, a wide range of jet observables have been developed for the study of the quark-gluon plasma (QGP) at RHIC and the LHC~\cite{Connors:2017ptx,Andrews:2018jcm}. Jet measurements yield a better proxy to parton kinematics than hadrons and are easier to interpret because they avoid the need for fragmentation functions. Moreover, modern jet substructure techniques offer new methods to explore QCD dynamics and control non-perturbative effects~\cite{Larkoski:2017jix,Asquith:2018igt}. 

The future EIC will provide a very clean environment where the underlying event and pileup are not significant, unlike hadronic collisions. This will allow for precise quantitative comparisons to perturbative QCD calculations. Moreover, the future EIC will allow for novel studies of hadronization, which can be investigated using various jet related observables.

Jet studies at the future EIC have been proposed to measure unpolarized and polarized parton distribution functions (PDFs) of the photon, proton and nuclei~\cite{Guzey:2019kik,Boughezal:2018azh,Klasen:2018gtb,Klasen:2017kwb,Chu:2017mnm}, along with the gluon polarization~\cite{Dumitru:2018kuw,Hinderer:2017ntk,Hinderer:2015hra}, nucleon transverse momentum dependent (TMD) PDFs~\cite{Liu:2018trl,Kishore:2019fzb,DAlesio:2019qpk,Zheng:2018awe,Boer:2016fqd,Dumitru:2015gaa,Pisano:2013cya,Boer:2010zf,Kang:2011jw}, generalized parton distributions~\cite{Hatta:2019ixj,Mantysaari:2019csc}, gluon saturation~\cite{Roy:2019hwr,Salazar:2019ncp,Dumitru:2016jku,Hatta:2016dxp,Altinoluk:2015dpi} and fragmentation in nuclei~\cite{Liu:2018trl,Sievert:2018imd,Kang:2013nha,Kang:2012zr}. We focus on tagged jets as precision probes of the nucleus via electron-jet correlations, which has recently been described in Ref.~\cite{Liu:2018trl}.

Despite the success of QCD in describing the strong interaction, the physics of parton interactions with QCD matter is not fully understood, as not everything can be calculated perturbatively. This is true both for the ``hot'' QCD matter produced in high energy nucleus-nucleus collisions, and the ``cold'' QCD matter probed via jet production in pp, p-A, p-p and e-A collisions~\cite{Accardi:2009qv}. Consequently, much of the theoretical work over the last two decades on the QGP provides a basis to build upon at the future EIC, which will unleash the precision era of QCD in nuclei. 

Naturally, the experiments at HERA--the first and only electron-proton collider--stand as a reference for future EIC jet measurements. We propose an approach different from that used for most jet measurements at HERA. Focusing on electroproduction in DIS, this work  also differs from recent work by Aschenauer et al.~\cite{Aschenauer:2019uex,Page:2019gbf} that focuses on jet photoproduction and gluon-initiated processes in e-p collisions.  

We study DIS jet production, $e A\to e'+\mathrm{jet}+X$ for event-by-event control of the kinematics ($x, Q^{2}$) that constrain the struck-quark momentum. We refer to this approach as electron-jet ``tag and probe'' studies. We identify several physics goals and identify approaches to realize them.

The paper is organized as follows: in Section~\ref{sec:requirementsTagandProbe} we describe the requirements and some experimental implications of the ``tag and probe'' measurements with electron-jet correlations; in Section~\ref{sec:Simulations} we describe the \textsc{Pythia8} simulation and the basic kinematic distributions of jet production; in Section~\ref{sec:Observables} we describe key observables with projected rates; in Section~\ref{sec:experimentalaspects} we discuss implications for future EIC detectors; and we conclude in Section~\ref{sec:conclusions}. 

\section{Requirements for Tag and Probe studies} \label{sec:requirementsTagandProbe}
In heavy-ion collisions, jets serve as ``auto-generated'' probes because they are produced in initial partonic hard scatterings prior to the formation of the QGP. As with any probe, its power relies on its calibration. In hadronic collisions, nature provides ``auto-calibrated'' processes such as $\gamma$-jet and $Z$-jet production. The mean free path of electroweak bosons in QCD matter is large whereas the jet interacts strongly, so coincidence measurements are a powerful way to constrain kinematics and systematically explore jet quenching in the QGP~\cite{Wang:1996yh}. 

\begin{figure}
    
    \centering
    \includegraphics[width=0.99\columnwidth]{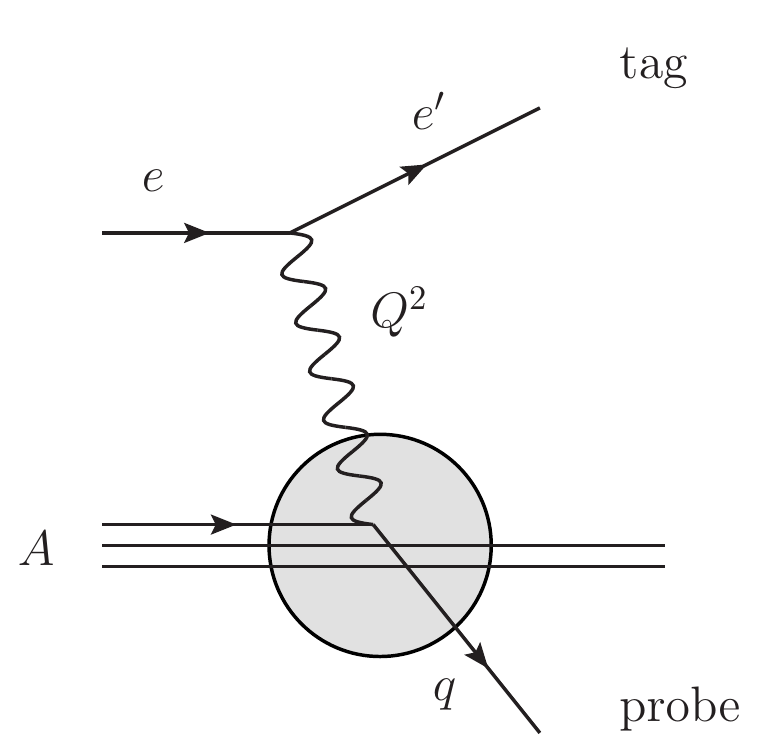}
    \caption{Leading-order DIS diagram. The scattered electron tags the kinematics of the struck quark, which then propagates through the nucleus before fragmenting into a jet of hadrons. The jet can thus be considered as a calibrated probe of the nucleus.}
\label{fig:DISdiagrams}
\end{figure}

Analogously, the virtual photon and the struck quark balance in DIS at leading order ($eq\to e'q$). We propose to use this process as a ``tag and probe'' to study the quark-nucleus interactions, as illustrated in Figure~\ref{fig:DISdiagrams} for a proton target. This approach differs from inclusive DIS, where the electron is considered the probe--our probe is the struck quark instead. Its color charge makes it suitable to study QCD in nuclei.

Unlike hadronic collisions, the electron is a fundamental particle and carries no color charge which simplifies the theory and provides a cleaner experimental environment suitable for accurate jet measurements. DIS offers a nearly pure quark-jet sample with little background from the underlying event. The nucleus has a high density of gluons at low temperature, which do not become highly excited in the collision. Consequently, the most challenging aspects of studying parton-QCD matter interactions in heavy ion or proton-nucleus collisions do not apply in this case. 

The basic requirements for ``tag and probe'' studies include:
\begin{enumerate}
    \item Kinematics such that the leading-order DIS 
    process dominates. 
    \item Event kinematics constrained by the electron measurement only.
    \item The jet must be matched to the struck quark by separating it from the beam remnant. 
\end{enumerate}
 
\noindent We explore the implications of each of these requirements in turn.

Initially, satisfying requirement 1 may appear straightforward. After all, the leading-order (LO) DIS diagram $(\gamma^{*}q\to\gamma^{*}q)$ is a pure electroweak process, whereas the higher-order DIS processes such as photon-gluon fusion $(\gamma^{*}g\to qq)$ or gluon bremsstrahlung $(\gamma^{*}q\to qg )$ are suppressed by $\alpha_{s}$. However, almost all jet studies at HERA suppressed the LO process by using the Breit Frame, in which the $\gamma^{*}$ points toward the positive $z$-direction with 3-momentum magnitude $Q$. At LO DIS, the struck quark flips its momentum from an incoming $-Q/2$ to $+Q/2$ in the $z$-direction, which is why the Breit frame is known as the ``brick-wall frame''. The LO DIS process produces a jet with zero transverse momentum, \pTjet, in the Breit frame, modulo the intrinsic transverse momentum of the quarks and the gluon radiation. Due to higher-order emissions, jets can pass that selection because multiple jets can balance each other's \pTjet~with respect to the $\gamma^{*}$ direction. The typical requirement of \pTjet> 4 GeV/c used at HERA~\cite{Newman:2013ada} effectively suppresses the LO DIS contribution, which was called ``{\it Quark-Parton Model background}'', and provides sensitivity to the gluon PDF and the strong coupling constant $\alpha_s$~\cite{Newman:2013ada}. 

The choice of reference frame is not a trivial one; one cannot simply transform the results presented in the Breit frame for the jet cross sections to another frame because of the minimum \pTjet~cut typically imposed. This cut ensures that theoretical calculations that require a scale related to the jet itself in addition to the $Q^{2}$ of the event is large enough for perturbative calculations to converge. 

In this work we show that jets with low \pTjet~in the Breit Frame are not only measurable and calculable, but offer a crucial tool at the future EIC. Instead of the Breit frame, we present results in the laboratory frame. Recent work by Liu et al.~\cite{Liu:2018trl} showed that the use of the e-A center-of-mass (CM) reference frame, which is related to the lab frame by a simple rapidity boost in the beam direction, provides a clear way to connect e-A results to hadron colliders. See also Ref.~\cite{Jager:2003vy}. 

We address requirement 1 by analyzing jets in the laboratory frame which is dominated by the LO DIS process.
Higher-order DIS processes are still present, but they can be taken into account by using e-p collisions as a baseline when studying e-A collisions. Moreover, NNLO calculations show that the contribution from photon and gluon-initiated processes are at the level of a few percent for {$Q^{2}> 25~ \GeVsquare$}~\cite{Abelof:2016pby,Frank}. 

Considering requirement 2, we note that the measurement of the scattered electron defines inclusive DIS and thus will likely drive the design of the future EIC detectors~\cite{EICRandD}. However, the energy and angular resolution of the scattered electron translates to a relative resolution of $x$ with a prefactor equal to the inverse of the event inelasticity 1/$y$; this follows from the definition\footnote{Inelasticity is bounded in the range between zero and unity, and in the nucleon rest frame corresponds to the fractional energy loss of the incoming electron.} of inelasticity $y = Q^{2}/sx$. Consequently, the resolution in $x$ diverges as the inelasticity goes to zero, $y\to0$. The limitation of the ``electron method'' to constrain $x$ and $Q^{2}$ was bypassed at HERA by using methods that rely on the hadronic final state~\cite{Newman:2013ada}, such as the Jacquet-Blondel method. These methods constrain the event kinematics by combining the information of all final-state particles in the event except the scattered electron. 

Using the Jacquet-Blondel method would not work for ``tag and probe'' studies, as it would amount to calibrating the jet probes with themselves. Consequently, the need to determine the kinematics purely from the scattered electron limits the ability to use low inelasticity events. Given detector response projections such as those presented in Ref.~\cite{EICRandD}, we note that even in the case of electron measurements with a combination of tracker and crystal calorimeter (with zero constant term and 2$\%$ stochastic term for $\eta<-2$) the resulting resolution in $x$ deteriorates rapidly for values of inelasticity $y<0.1$. We therefore conclude that the tag and probe method requires events with inelasticity $y>0.1$. The exact value of the inelasticity $y$ cut can be optimized based upon the actual detector performance.

We identify the kinematic selection criteria needed to meet requirement 3 and present the results in Section~\ref{sec:beamremnant} after we introduce our simulations and show the kinematic distributions of jets expected at the future EIC in the next section.

\section{Simulations} \label{sec:Simulations}

We use \textsc{Pythia8}~\cite{Sjostrand:2007gs} to generate neutral-current DIS events in e-p collisions with energies of 20 GeV for the initial state electron and 100 GeV for the proton, resulting in a center-of-mass energy of $\sqrt{s}=89$~GeV. While proton beam energies of up to 250~GeV are considered in the future EIC designs~\cite{Aschenauer:2014cki,Satogata:2018dmt}, the per-nucleon energy of the nuclear beams is reduced by a factor of $Z/A$, which is $\approx 0.4$ for heavy nuclei.

We select particles with $\pT>250$ MeV/c and $|\eta|<4.5$ in the lab frame\footnote{We follow the HERA convention to define the coordinate system we use throughout this paper. The $z$ direction is defined along the beam axis and the electron beam goes towards negative $z$. The pseudorapidity is defined as $\eta = -\ln[\tan(\theta/2)]$, where the polar angle $\theta$ is defined with respect to the proton (ion) direction.}, excluding neutrinos and the scattered electron (which we identify as the highest \pTelectron~electron in the event). The asymmetry of the beam energies creates a boost of the e-A center-of-mass frame relative to the laboratory frame given by $\eta^{\mathrm{lab}} = \etacm + 0.5\ln(E_{p}/E_{e}) = \etacm + 0.80$ for the kinematics considered here.

We use the \textsc{Fastjet}3.3 package~\cite{Cacciari:2011ma} to reconstruct jets with the anti-$k_{\mathrm{T}}$ algorithm~\cite{Cacciari:2008gp} and $R=1.0$. For most studies, we use the standard recombination scheme (``E-scheme''), where the jet clustering just combines 4-vectors, but we also present some results with the ``winner-take-all'' scheme~\cite{Salam:WTAUnpublished,Bertolini:2013iqa} where the jet axis is aligned with the more energetic branch in each clustering step.

Our choice of the distance parameter $R=1.0$ follows the HERA experiments where it was found that this large value reduces hadronization corrections for inclusive jet spectra to the percent level~\cite{Newman:2013ada}. At the future EIC, smaller $R$ values might help to tame power corrections for jet substructure observables which we leave for future work, see also Ref.~\cite{Aschenauer:2019uex}. 

\textsc{Pythia8} uses leading-order matrix elements matched with the showering algorithm and the subsequent hadronization. For DIS, \textsc{Pythia8} relies on the \textsc{DIRE} dipole shower~\cite{Hoche:2015sya} to generate high order emissions. Our simulations do not include QED radiative corrections or detector response. Initial and final-state QED radiative corrections ``smear'' the extracted $x$ or $Q^{2}$ from the measured electron angle and momentum with respect to the Born-level values. We select observables that minimize the sensitivity to radiative corrections, and further reduce radiative effects in three ways: require inelasticity $y<0.85$, which removes the most sensitive phase space; construct ratios of cross sections (semi-inclusive DIS jet cross sections and inclusive DIS cross section); and bin in \pTelectron. The \pTelectron~variable is insensitive to initial-state QED radiation and has reduced sensitivity to collinear final-state radiation. Moreover, ratios between measurements in e-A and e-p data will further suppress the impact of radiative corrections. 

We use the EPPS16 nuclear PDFs~\cite{Eskola:2016oht} for the $Pb$ nucleus, to approximate hard scatterings in e-A collisions in our e-p sample. Of course, the underlying event in e-A is not simulated in this approach. However, due to the absence of multi-parton interactions in DIS, the underlying event is expected to be small compared to p-A collisions. We do not include the impact of Fermi motion in our simulations which is only relevant for the very high-$x$ region.

We require $Q^{2} > 1$ \GeVsquare, the invariant mass of the hadronic final state $W^{2}> 10$ \GeVsquare, and the inelasticity of the event of $0.1<y<0.85$. The lower elasticity limit avoids the region where one cannot constrain the event kinematics with the electron (as discussed in Section~\ref{sec:requirementsTagandProbe}), whereas the upper limit avoids the phase space in which QED radiative corrections are large.

We do not simulate photoproduction processes which are defined~\footnote{There is a continuum between the photoproduction region $Q^{2}\approx 0$ and electroproduction region at larger $Q^{2}$. The dividing line is arbitrary, but it is typically defined as $Q^{2}=1$ GeV$^{2}$~\cite{Accardi:2012qut}, which we adopt for our studies.} by $Q^{2}\approx 0$. The photoproduction process is similar to jet production in hadron collisions, which includes all the complications we aim to avoid as well as sensitivity the relatively poorly known photon PDFs. Therefore, photoproduction of jets is a background for this study, and can be reduced to a negligible level by requiring large values of $Q^{2}$~\cite{Abelof:2016pby,Frank}.

We simulate $10^7$ events to ensure the statistical precision of the Monte Carlo simulation. The projected rates correspond to an integrated luminosity of 10 fb$^{-1}$, which can be collected in a few months of e-p  running. While the cross sections for hard processes in e-A are higher by a factor of $A$, the luminosity expected for ions is smaller approximately by a factor of $A$, leading to similar rates for e-A and e-p collisions at the future EIC. 

\subsection{Differential cross section and event kinematics}
\begin{figure}
    \centering
    \includegraphics[width=0.49\textwidth]{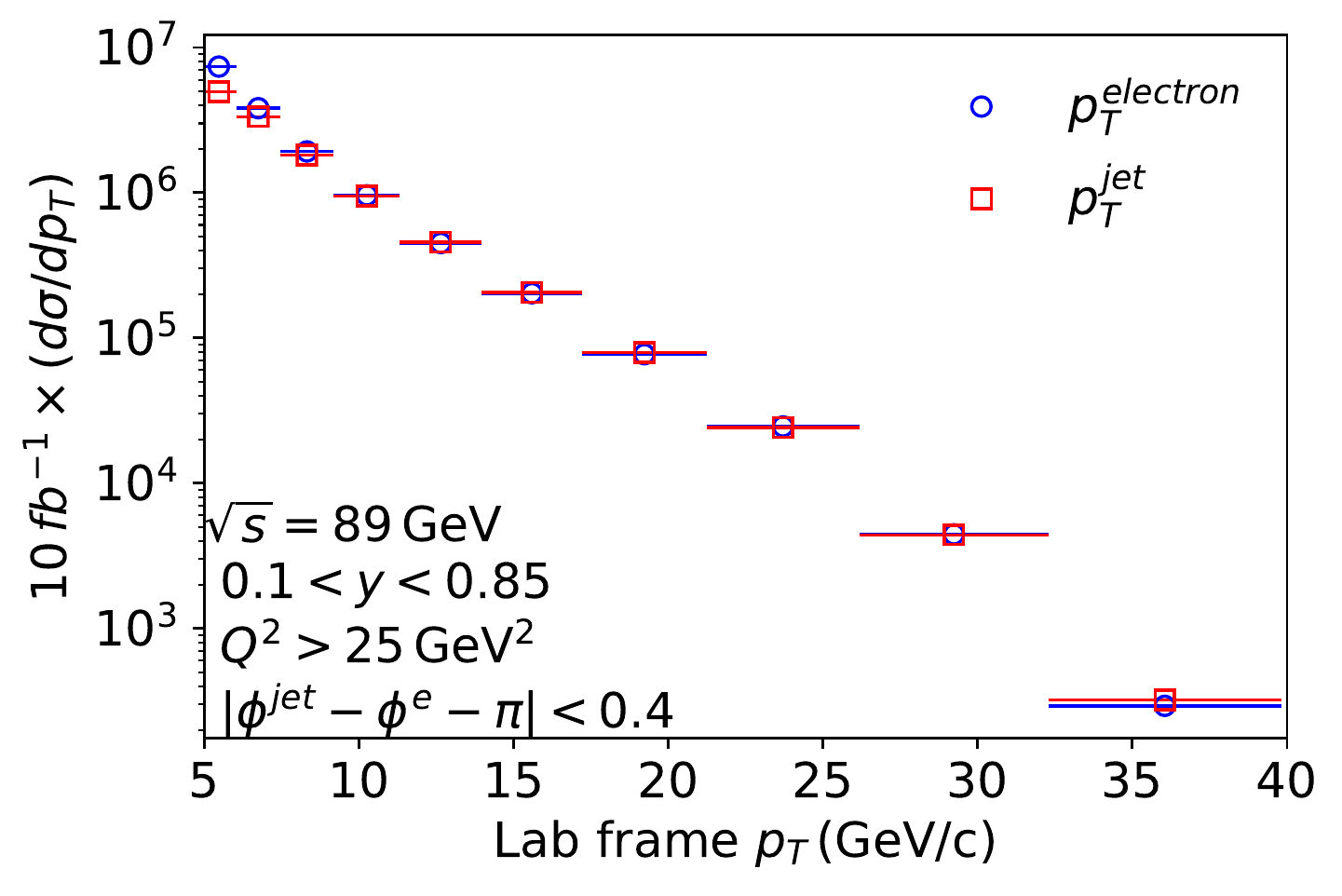}
\caption{Yield of electrons and jets for 10 fb$^{-1}$. The \pT~here is defined in the laboratory frame (or equivalently in the electron-nucleon center-of-mass frame). The jets are reconstructed with the anti-$k_{\mathrm{T}}$ algorithm with $R=1.0$. The projected statistical uncertainty is negligible for most of the kinematic region and smaller than the marker size.}
\label{fig:spectra}
\end{figure}

Figure~\ref{fig:spectra} shows the expected yield of electrons and jets for 10 fb$^{-1}$ integrated luminosity, as a function of \pT~in the lab frame. The \pT~in the lab frame is equivalent to the \pT~in the electron-nucleon center-of-mass frame as it is invariant under boosts in the longitudinal direction. In addition, we apply a cut on the azimuthal angle between the electron and the jet $|\phi^{jet}-\phi^e-\pi|<0.4$, which suppresses jets arising from the fragmentation of the beam remnant as we will show in Section~\ref{sec:beamremnant}.

The transverse momentum spectra reach up to $\pT\approx 35$~GeV/c. The electron and jet distributions generally agree well since only a single jet is produced in DIS. This is not the case at low \pT, where $\alpha_s$ is larger and parton branching processes/out-of-jet emissions generate low $p_{\rm T}^{jet}$ jets that do not pass the selection criteria. In addition, hadronization effects become more important at low $p_{\rm T}^{jet}$. 

Collecting 10 fb$^{-1}$ of data would yield  statistical uncertainties at the sub-percent level. Of course, this depends on detector acceptance, efficiencies, and triggering. The high luminosity of the future future EIC will allow for a comparison of several different nuclei, along with detailed studies required to constrain systematic uncertainties. 

The electron transverse momentum and pseudorapidity are not variables commonly used to characterize the event kinematics in DIS, but they are closely related to $Q^{2}$ and $x$ by $Q^{2}= -\hat{t} =  \sqrt{s}\, \pT^{e} e^{-\eta^{e}}$ and $\hat{u} = x\sqrt{s}\,\pT^{e}e^{+\eta^{e}}$, where $\eta^{e}$ is the pseudorapidity of the electron in the electron-nucleon CM frame~\cite{Liu:2018trl} and $\hat{t}$ and $\hat{u}$ are the Mandelstam variables. 

\begin{figure}
    \centering
    \includegraphics[width=0.49\textwidth]{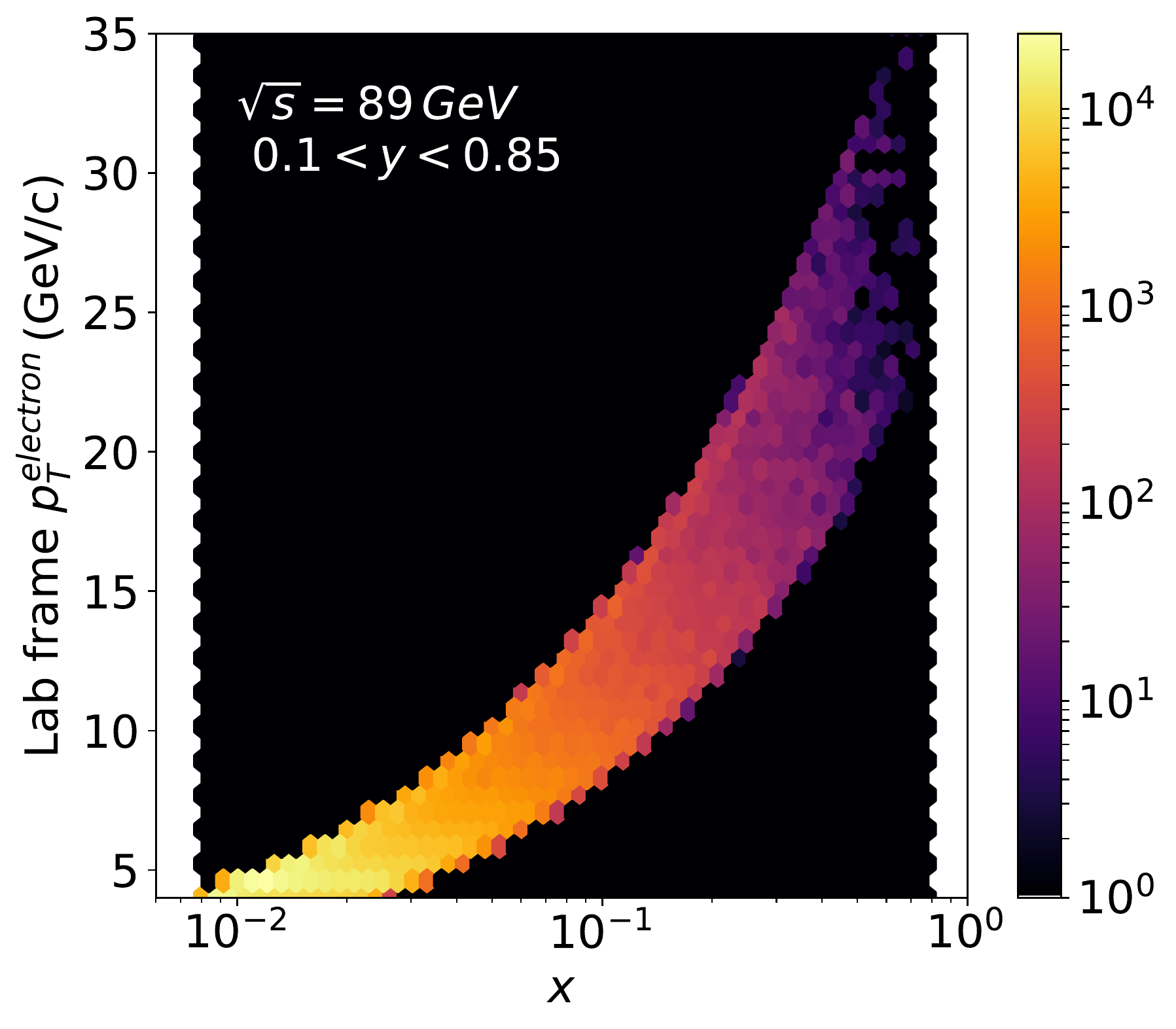}
\caption{Electron transverse momentum vs Bjorken $x$. The beam energies of the simulation are 20~GeV for the electron and 100~GeV for the proton. }
\label{fig:eptvsx}
\end{figure}

Figure~\ref{fig:eptvsx} shows \pTelectron~and $x$ distributions for events passing the cuts listed above. The observed ``strip'' is the result of the inelasticity selection. In particular, events with low $Q^{2}$/high $x$ yield low inelasticity ($y = Q^{2}/sx$), which is removed by our requirement $y>0.1$. Nevertheless, we obtain a wide coverage in $x$ with jets, spanning the shadowing, anti-shadowing and EMC regions in e-A collisions. While these regions have been studied before in inclusive DIS and semi-inclusive DIS in fixed-target experiments, the future EIC energies will allow the measurement of jets over a wide range of $Q^{2}$.

\subsection{Jet energy and pseudorapidity distributions}
\begin{figure}
    \centering
    \includegraphics[width=0.49\textwidth]{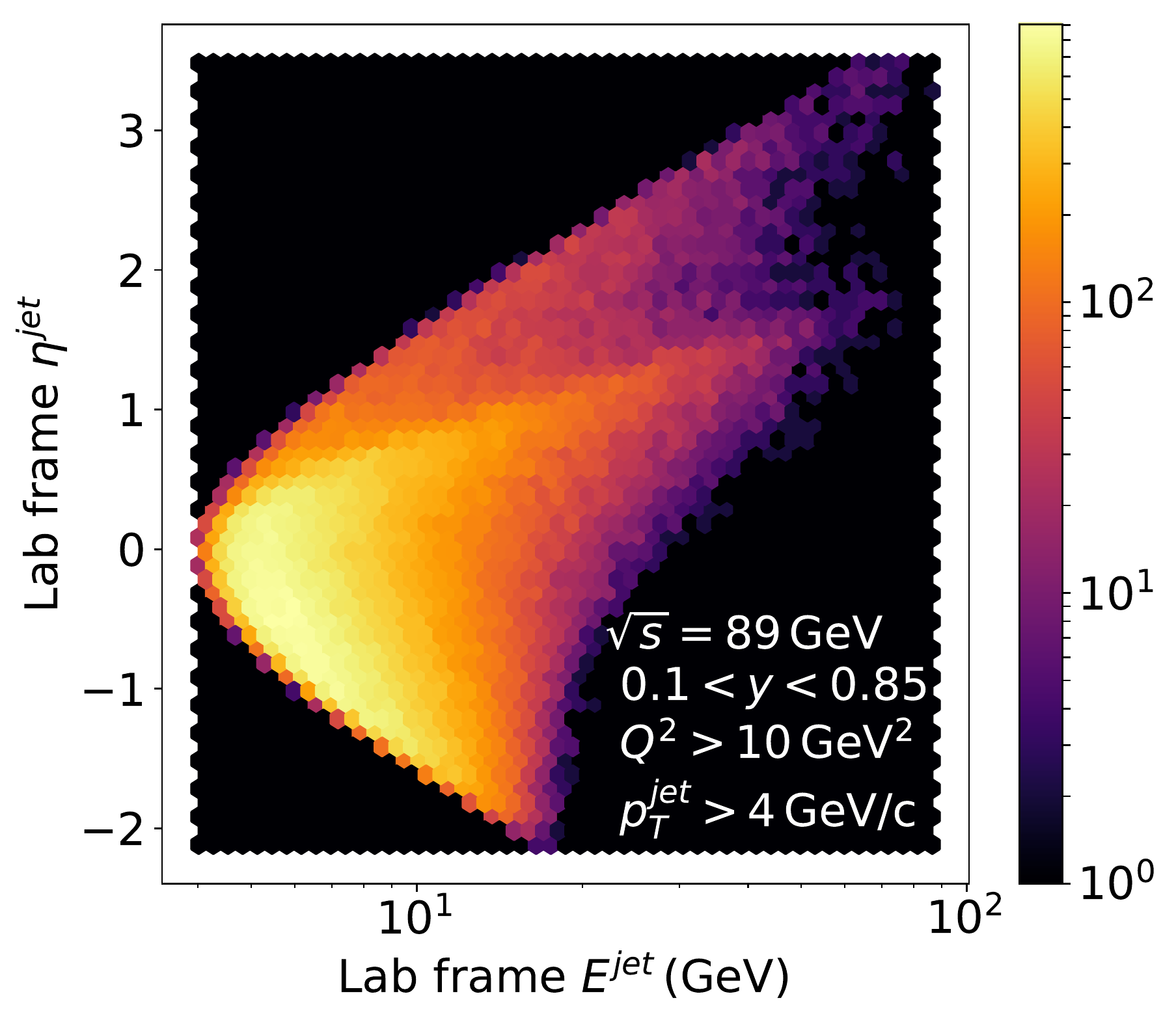}
\caption{Jet energy vs jet pseudorapidity (in the lab frame). $\etalab$ is defined as positive in the proton (ion)-going direction. The jets are defined with radius $R=1.0$ and the anti-$k_{\mathrm{T}}$ algorithm. The beam energies of the simulation are 20~GeV for the electron and 100~GeV for the proton. }
\label{fig:etavsenergy}
\end{figure}

Figure~\ref{fig:etavsenergy} shows the jet pseudorapidity and energy in the lab frame. The exact shape of the distribution is due to the inelasticity selection, the asymmetric nature of the collision, and the rapidity boost of $\Delta \eta \approx 0.8$ due to different beam energies. The jet energy at mid-rapidity ($\etalab\approx0$) is limited to $\approx$~30~GeV, whereas in the backward direction it reaches only about $\approx$~20~GeV, as it is limited by the electron beam energy. On the other hand, jets with energies in the range 50--100 GeV are produced in the forward direction ($\etalab>1.0$). 

\subsection{Number of jet constituents}

\begin{figure}
    \includegraphics[width=0.49\textwidth]{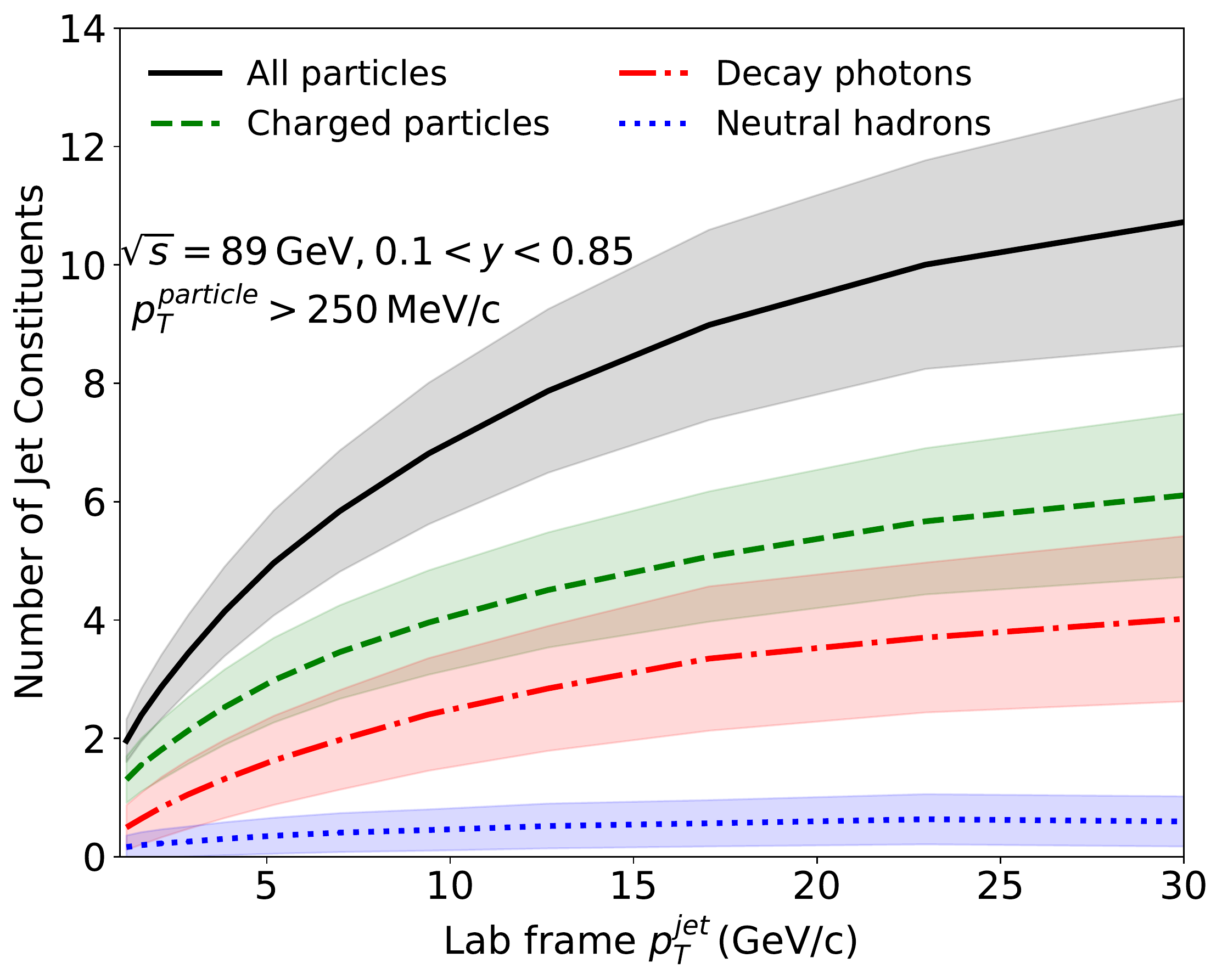}
    \centering
\caption{Number of particles inside the jets as a function of the transverse momentum \pTjet in the lab frame. The jets are defined with radius $R=1.0$ and the anti-$k_{\mathrm{T}}$ algorithm. The error bands represent the standard deviation of the distribution for each \pTjet~interval. The pseudorapidity range (in lab frame) of $|\etalab|<3.0$ is considered.}
\label{fig:numberofconstituents}
\end{figure}

Figure~\ref{fig:numberofconstituents} shows the number of particles in the jets as a function of \pTjet~for charged particles, photons from the decay of neutral mesons, and neutral hadrons. There is a gradual increase with \pTjet. We checked that there is no significant change with pseudorapidity of the jet within the range  $|\etalab|<3.0$. Therefore, the particle multiplicity does not depend on the jet energy, but only on its \pTjet. We also find no $Q^{2}$ dependence within 1--1000 GeV$^{2}$. 

While jet algorithms can in principle ``find'' jets with low transverse momentum which may contain only very few particles, the question is whether useful information can be extracted from these ``mini-jets''. The answer depends on the observable under consideration and requires a comparison to perturbative QCD calculations including QCD scale uncertainty estimates, which increase at low $\pTjet$. While a generic cut on particle multiplicity or transverse momentum is somewhat arbitrary, we follow here the precedents set by experiments at HERA and RHIC, and require \pTjet $\ge$ 4 GeV/c.

\begin{figure*}
    \centering  
    \subfigure[Kinematic distributions of the scattered electrons and struck quarks]{\includegraphics[width=0.32\linewidth]{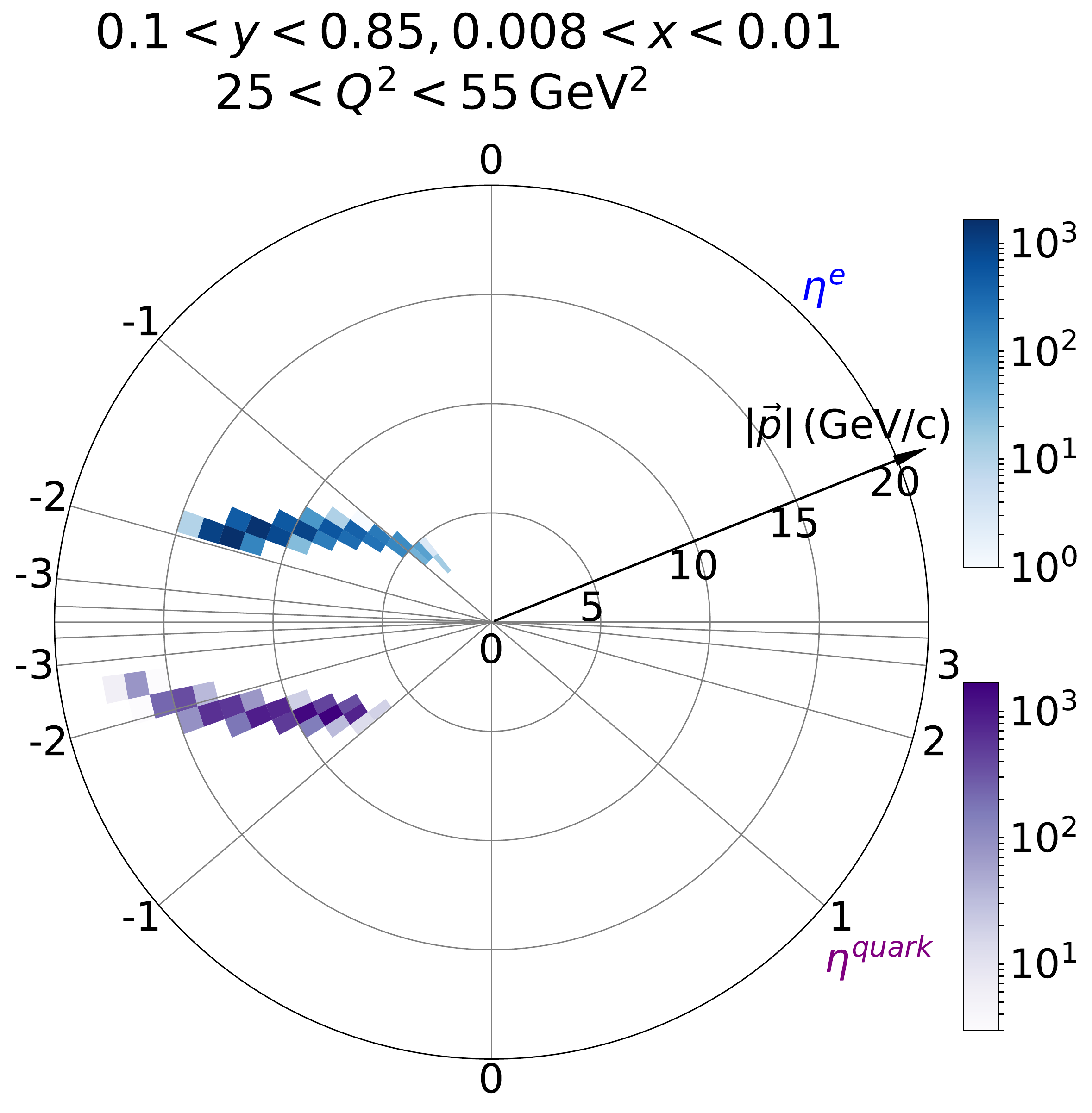}}
    \subfigure[Kinematic distributions of the scattered electrons and jets]{\includegraphics[width=0.32\linewidth]{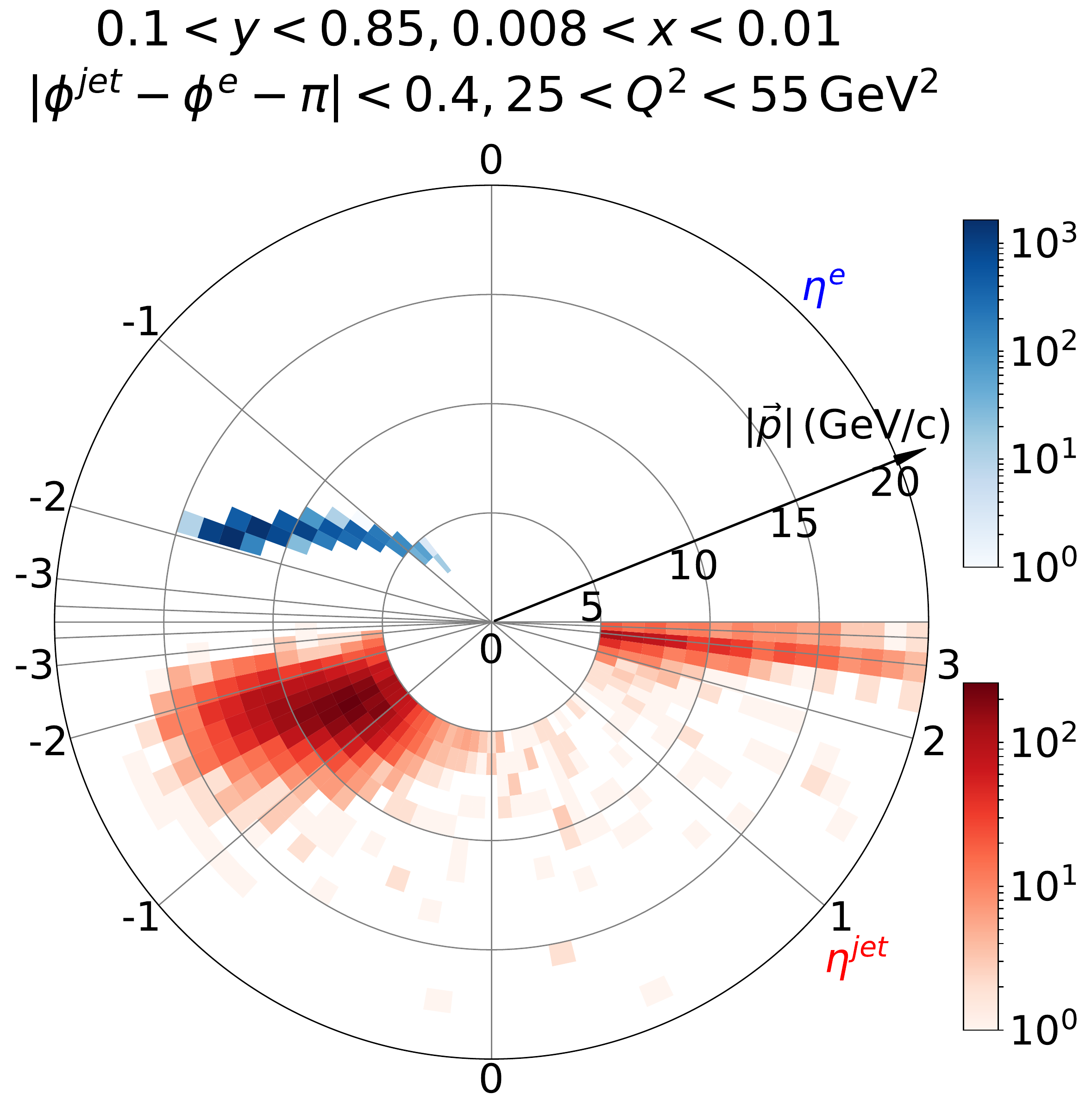}}
    \subfigure[Kinematic distributions of the scattered electrons and hadrons]{\includegraphics[width=0.32\linewidth]{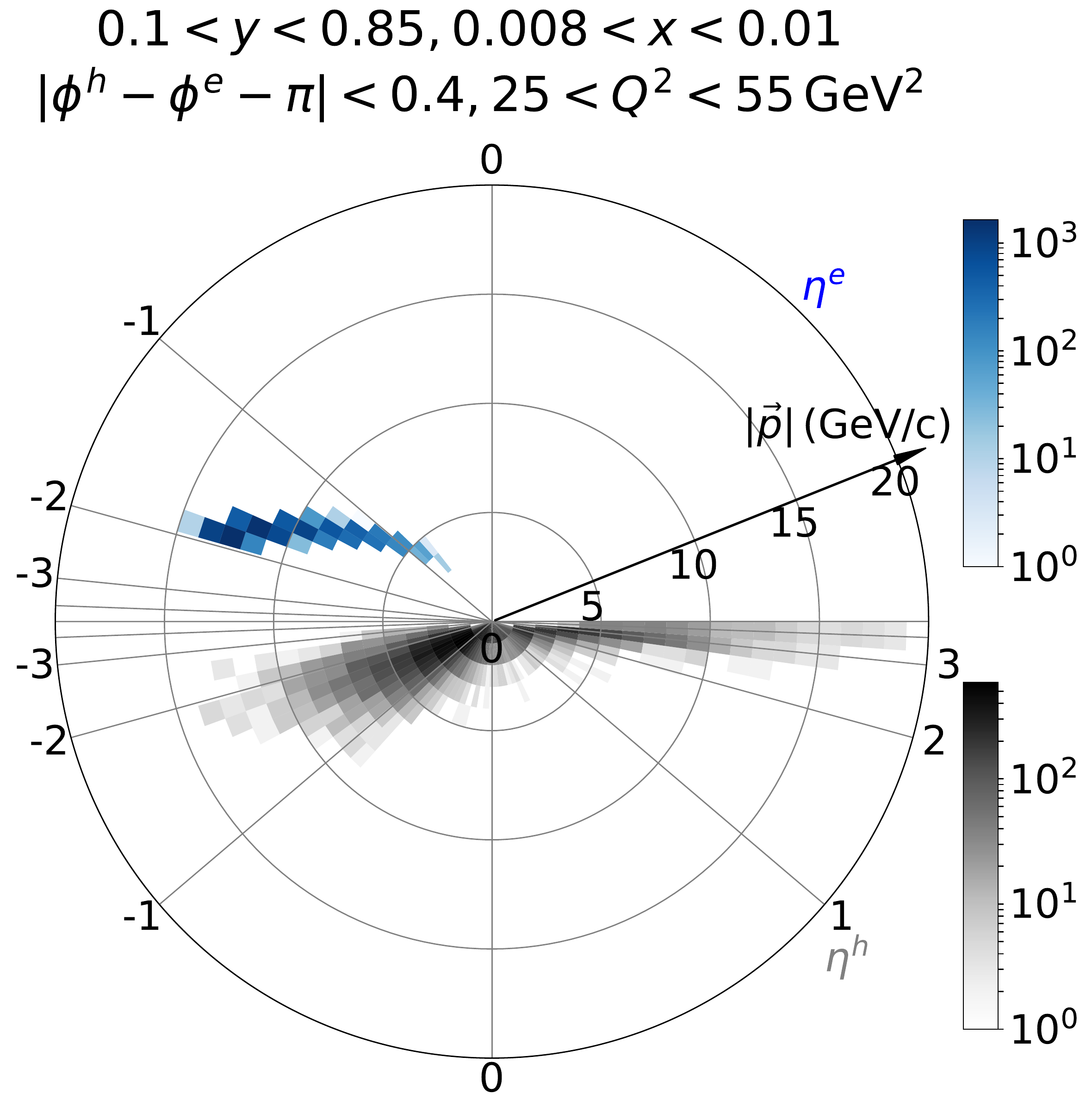}}
    \subfigure[Kinematic distributions of the scattered electrons and struck quarks]{\includegraphics[width=0.32\linewidth]{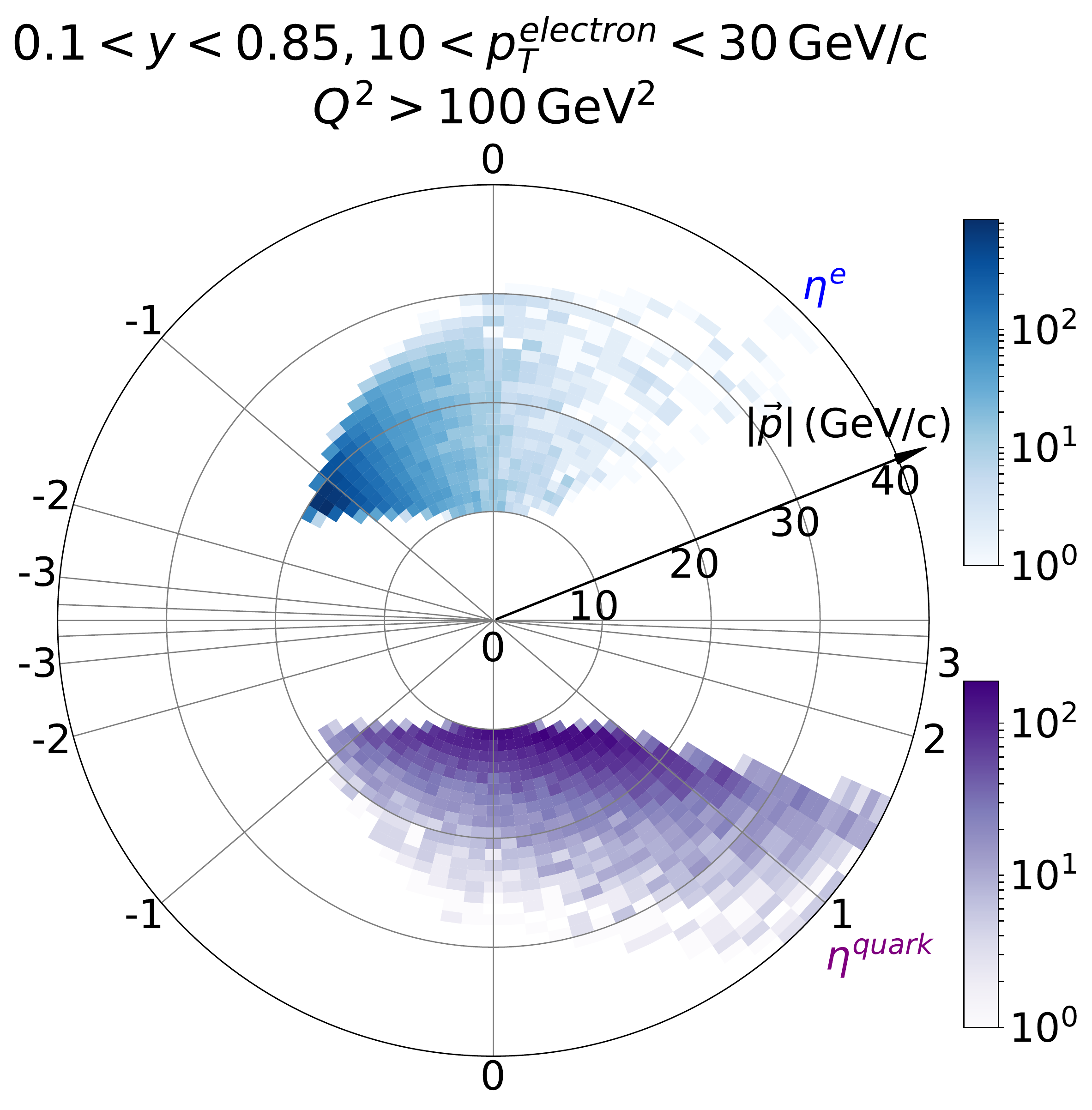}}
    \subfigure[Kinematic distributions of the scattered electrons and jets]{\includegraphics[width=0.32\linewidth]{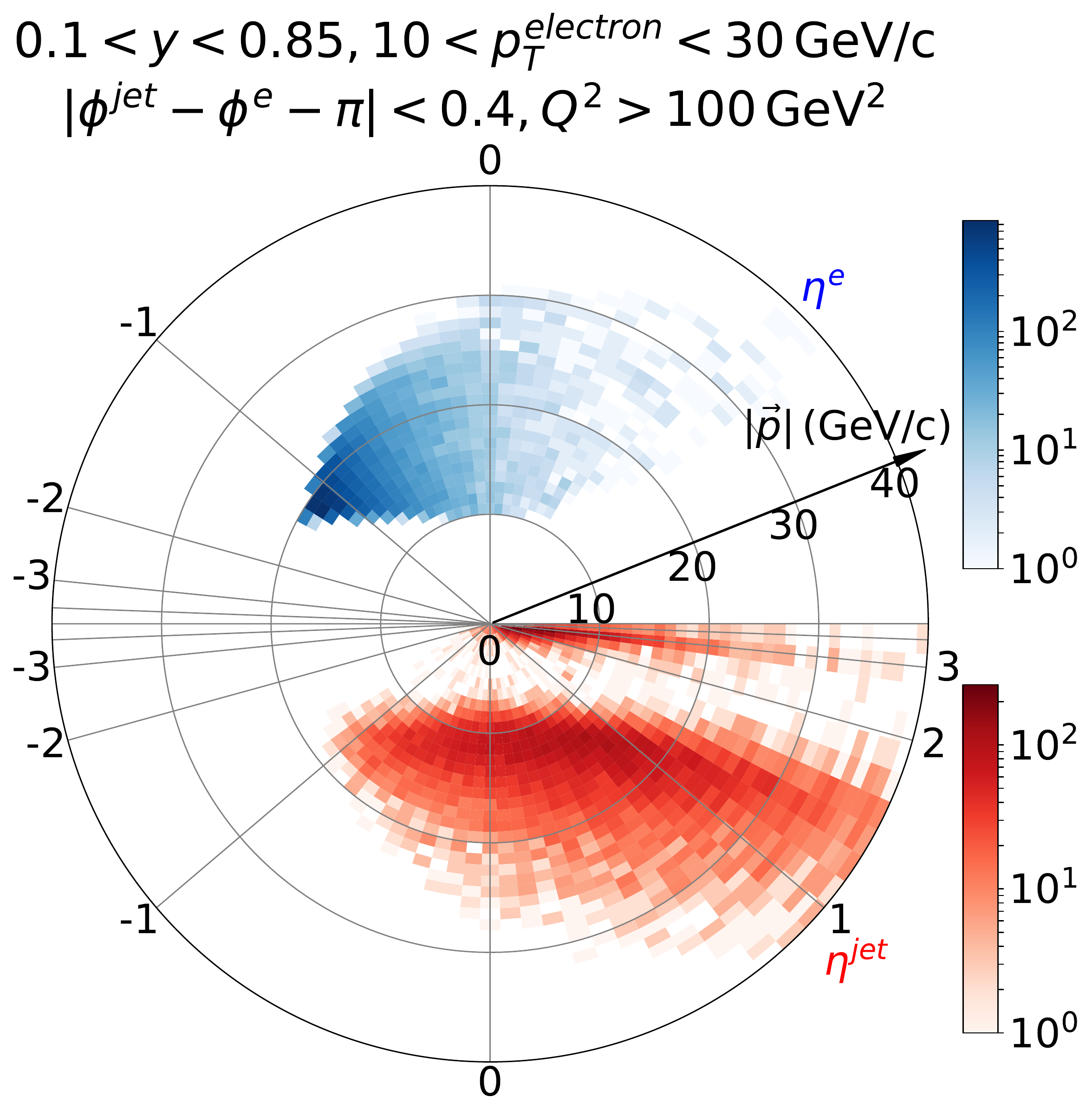}}
    \subfigure[Kinematic distributions of the scattered electrons and hadrons]{\includegraphics[width=0.32\linewidth]{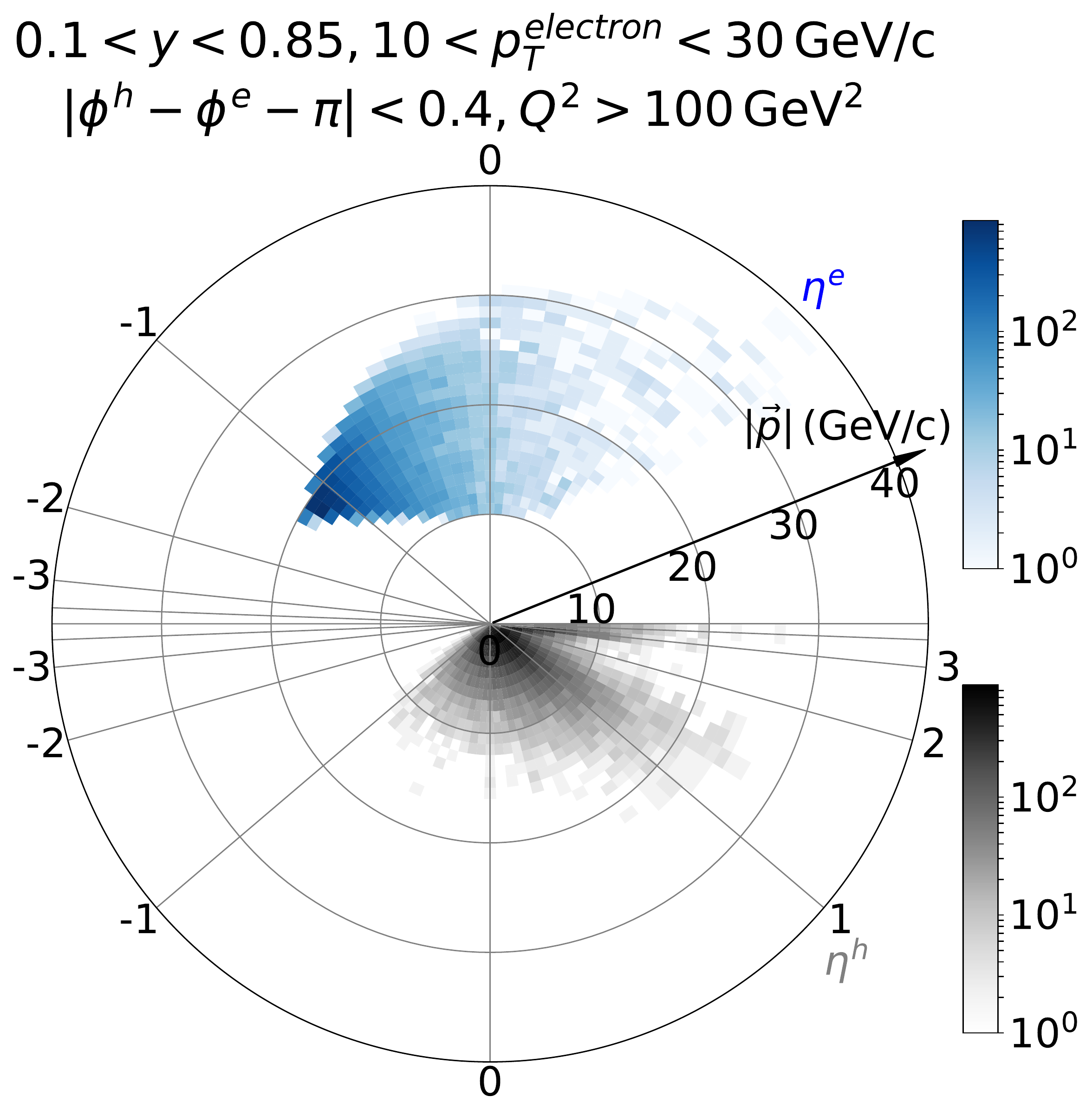}}
    \caption{Polar plots of the kinematic distributions of the particles and jets produced in DIS. The top half of each circle shows the pseudorapidity and 3-momentum of the scattered electron in the angular and radial direction, respectively. The bottom half of each circle shows the pseudorapidity and momentum of particles and jets: the polar plots of the struck quark are on the left, the jets in the middle, and particles on the right.}
    \label{fig:polarplot}
\end{figure*}
\subsection{Separation of struck-quark and beam-remnant fragmentation}
\label{sec:beamremnant}

As noted in requirement 3 in Section~\ref{sec:requirementsTagandProbe}, using the struck quark as a tagged probe requires kinematic cuts to select jets arising from that quark. One of the benefits of the collider mode is that beam remnants continue to move in the beam direction while the particles produced by the fragmentation of the struck quark might be separated. This picture is complicated by the process of hadronization. As noted by Aschenauer et al.~\cite{Aschenauer:2019kzf}, hadrons from beam-remnant and struck-quark fragmentation largely overlap in rapidity for all $Q^{2}$ accessible at the future EIC. 

The separation of struck-quark and beam-remnant fragmentation is central for theoretical studies to interpret the data, as relevant factorization theorems apply to the struck-quark fragmentation only\footnote{Here we use the terms struck-quark and beam-remnant fragmentation for clarity, which corresponds to current and target fragmentation that are also used in the literature.}. Recent theoretical studies have focused on this issue~\cite{Boglione:2016bph,Boglione:2019nwk}. In this work, we explore the beam-remnant separation in an empirical way by using the hadronization model in \textsc{Pythia8} and compare results using jets and hadrons. 

As an aid in identifying the struck-quark fragments, we construct polar plots tracking the scattered electron and struck quark as well as jets and hadrons. Examples are shown in Figure~\ref{fig:polarplot}. The top half of each circle shows the pseudorapidity and 3-momentum of the scattered electron in the angular and radial direction, respectively. The bottom half shows the rapidity and momentum of the hadronic partners.
Polar plots of the scattered electron and struck quark are shown on the left, jets in the middle, and hadrons on the right.

Figures~\ref{fig:polarplot}-a),b) and c) show where the reaction products go when the struck quark $x$ is low, from 0.008 to 0.01. As expected for DIS off quarks at low-$x$, the struck quark travels to negative rapidity, i.e in the electron-going direction. Figure~\ref{fig:polarplot}-b) shows two clear sources of jets: one corresponding to the struck quark and the other to the beam remnant.  The two jet sources are quite well separated in pseudorapidity, making a selection of the struck quark jet straightforward in this case. We found that a minimum of $Q^{2}>25$ GeV$^{2}$ is needed to achieve this clean separation for this kinematic interval; decreasing $Q^{2}$ leads to a worsening of the separation. Figure~\ref{fig:polarplot}-c) shows the distribution of single hadrons. While a correlation with the pseudorapidity of the parent quarks is present, it is significantly smeared for lower \pT~hadrons, making the experimental separation of struck quark and beam remnant products more difficult than with jets. The $|\phi^{jet} - \phi^e - \pi|<0.4$ cut in the middle and right plots requires the electron and jet to be back-to-back in azimuthal angle, as explained below.

This clear identification of the struck-quark at low-$x$ guarantees access to the dense gluon-dominated matter at small $x$ which requires selecting DIS off a parton which is itself at small $x$. This parton then transits the dense matter on its way to the detector. Comparing jets from such partons in scattering from different nuclei will allow us to quantify the transport properties of the dense matter. 

The bottom panels show a similar set of polar plots selecting $10<\pTelectron< 30$ GeV/c and $Q^{2}>100$ GeV$^{2}$. Figure~\ref{fig:polarplot}-d) shows that the scattered quarks start to go in the hadron beam-going direction, but they are still dominantly 
at pseudorapidities less than 2. Figure~\ref{fig:polarplot}-e) shows that the separation of the struck quark and beam remnant jets is also clearly feasible for these kinematics, even though the pseudorapidity separation is smaller. The smearing for single hadrons, however, is much larger, as visible on Figure~\ref{fig:polarplot}-f). 
For this electron \pT ~range, $Q^{2}>100~\text{GeV}^2$ is required to obtain the separation with jets; significantly lower $Q^{2}$ values lead to a much larger overlap.

We conclude that the prospect for separating the struck quark and beam remnants looks very promising with jets.
\section{Observables} \label{sec:Observables}

We now turn to jet observables of interest for probing properties of gluon-dominated matter in nucleons and nuclei. Sections~\ref{sec:ptbalance} and~\ref{sec:azimuthaldecorrelation} show the transverse momentum and azimuthal balance of the electron and jets; and section~\ref{sec:groomedjets} describes the groomed jet radius. 

\subsection{Transverse momentum balance}
\label{sec:ptbalance}

\begin{figure}
    \centering
    \includegraphics[width=0.45\textwidth]{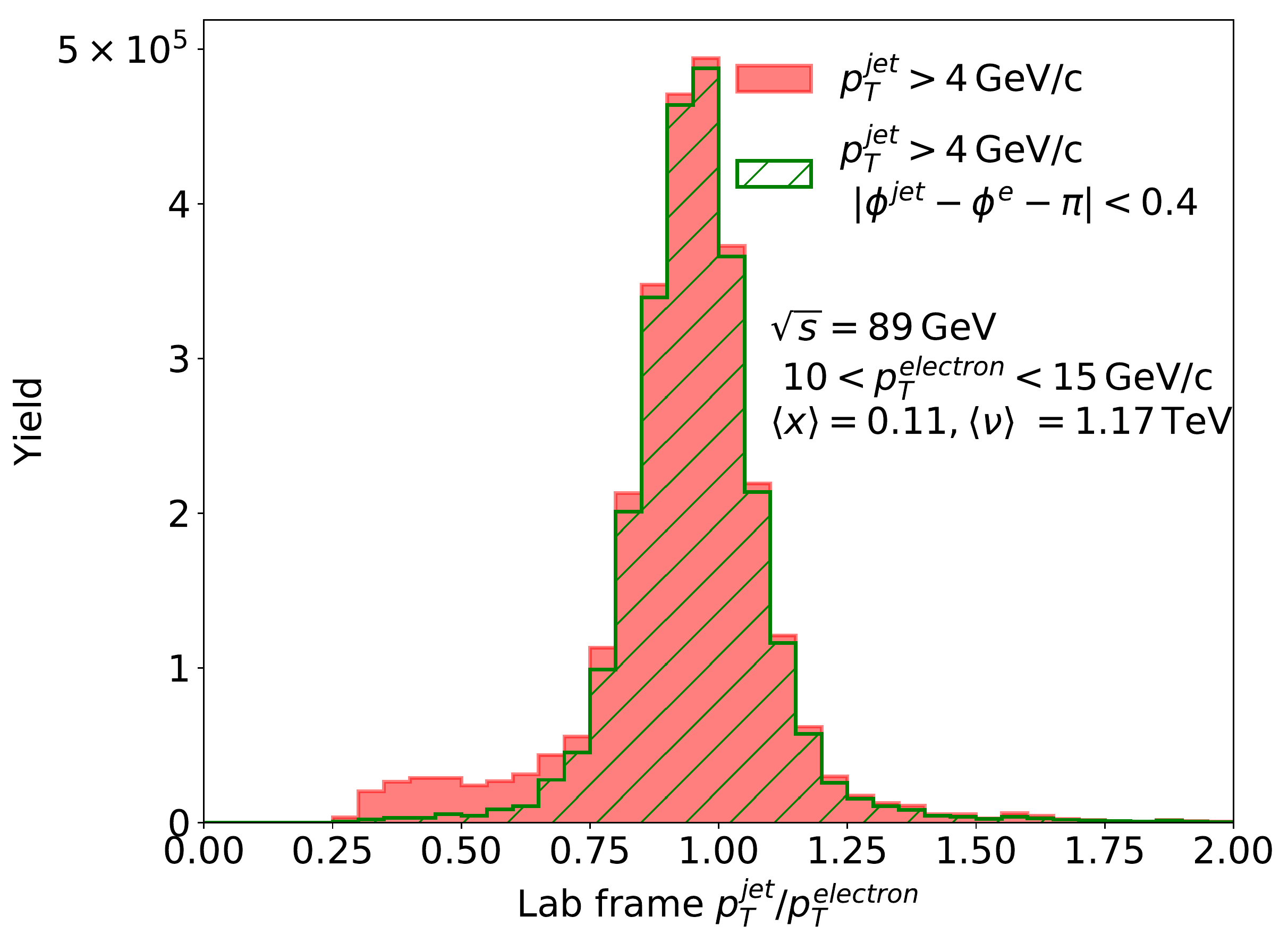}
\caption{Transverse momentum balance between the scattered electron and jets in DIS events. The red (filled) distribution shows all electron-jet pairs, whereas the green (shaded) includes a selection on the azimuthal difference, which is expected at leading-order DIS. The projected statistical uncertainties are negligible and not shown.}
\label{fig:ptbalance}
\end{figure}

A key measurement sensitive to the mechanism of quark energy loss in the nucleus is the ratio of the electron to jet transverse momentum, since
the electron tags the struck-quark \pT. Figure~\ref{fig:ptbalance} shows the transverse momentum balance between the scattered electron and jet for $10< \pTelectron < 15$ GeV/c and $\pTjet > 4$ GeV/c. The distribution peaks around unity as expected for DIS. The width of the distribution arises from initial state radiation, out-of-jet emissions and hadronization~\cite{deFlorian:2010vy}. Applying a cut on the azimuthal difference between the scattered electron and jets $|\phi^{jet} - \phi^e - \pi| <0.4$ suppresses low-\pTjet~jets not associated with the scattered electron, i.e jets from beam remnant fragmentation. 

An important variable for these studies is $\nu$, which is the the virtual photon energy (struck-quark energy) in the rest frame of the nucleon and is given by $\nu = Q^{2}/2mx$ with $m$ the nucleon mass\footnote{The variable $\nu$ plays a central role in characterizing quark-nucleus interactions, which is why it has been used by all previous fixed-target e-A scattering experiments~\cite{Accardi:2009qv}; it has also been recognized as a key variable for studies of hadronization at the future EIC because it controls Lorentz dilation in the rest frame of the nucleus and therefore dictates whether hadronization occurs inside or outside the nucleus~\cite{Accardi:2012qut}.}. For this kinematic selection, 
the average $x$ is 0.11 and the average $\nu$ is 1.1 TeV. The same $x$ region is accessible in fixed-target experiments, for example those ongoing at the Jefferson Laboratory CEBAF, but with $\nu$ values of only a few GeV (or equivalently, low $Q^{2}$). This illustrates that future EIC experiments will explore kinematics that represent {\it terra incognita} even in ``known'' $x$ regions. In particular, we would be able to answer ``how does the nucleus react to a fast moving quark'' at the TeV scale, whereas all previous fixed-target experiments reached $\nu$ values of ${\cal O}(10~\text{GeV})$. Given the large number of events expected at the future EIC, 
it will be possible to bin finely in either $x$ or $\nu$, once radiative corrections are applied. 

At the future EIC we will be able to explore in detail the kinematic dependence of the jet transport coefficient, $\qhat$, where $\hat q L$ describes the typical transverse momentum squared acquired by a parton traversing the medium of length $L$. The kinematic dependence of $\hat q$ in cold nuclear matter is under active investigation, see for example recent work in Refs.~\cite{Kumar:2019uvu,Ru:2019qvz,Zhang:2019toi}. The kinematic coverage of future EIC semi-inclusive DIS data (hadron and jet) will be several orders of magnitude larger than the existing semi-inclusive data and will be much more precise; therefore, it will allow for definitive conclusions on the properties of the jet transport coefficient $\hat q$. In general, these results may also illuminate studies of the QGP in heavy-ion collisions.

Energy loss studies at the future EIC will provide a more accurate measurement of \qhat~in nuclei than is likely to be achieved in p-A collisions. There are several reasons: DIS in e-A has much less background than the underlying event in p-A collisions; DIS provides an almost pure quark probe instead of quark-gluon fractions that depend on kinematics; in DIS a virtual photon interacts with the quark, experiencing no initial state scattering and leaving a medium that is static and not affected by QCD multi-parton interactions; event-by-event tagging of the struck quark in DIS improves the precision of the measurement and theoretical calculations; and the future EIC luminosity will offer superb statistics. 

\subsection{Electron-jet azimuthal correlation~\label{sec:azimuthaldecorrelation}}

\begin{figure}
    \centering
    \includegraphics[width=0.45\textwidth]{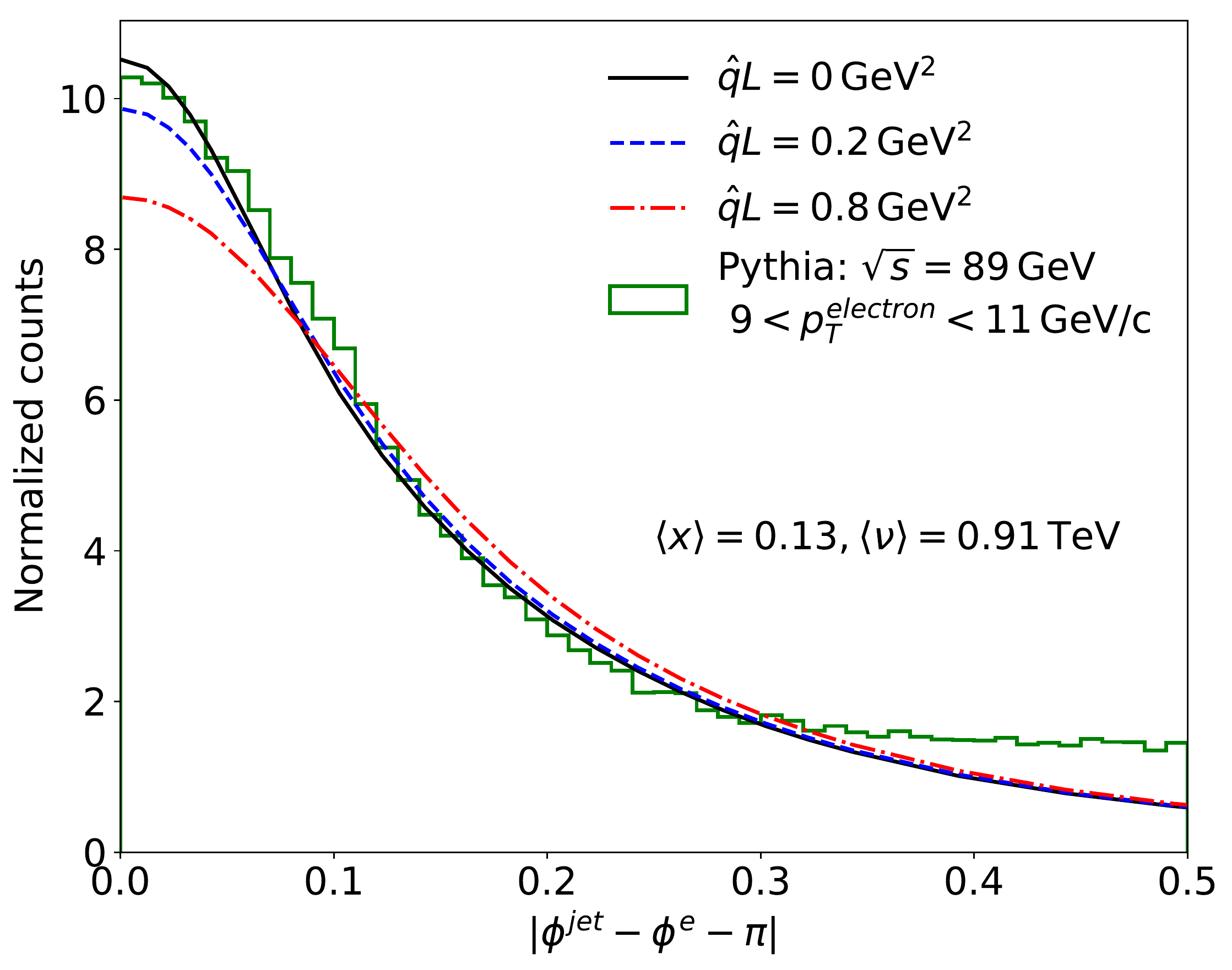}
\caption{The azimuthal angle correlation between the electron and jets in DIS events. The azimuthal angle is defined in the electron-nucleon frame. The theoretical calculations by Liu et al.~\cite{Liu:2018trl} are shown in the vacuum (solid black curve) and including medium effects (dashed curves) for typical values of $\hat qL$. All distributions are normalized to unity. The results presented in this Figure do not contain an inelasticity cut for consistency with~\cite{Liu:2018trl}. The projected statistical uncertainties are negligible and not shown.} 
\label{fig:azimuthal}
\end{figure}

Figure~\ref{fig:azimuthal} shows \textsc{Pythia8} results for the azimuthal difference $|\phi^{jet} - \phi^e-\pi|$ between the scattered electron and jets. The azimuthal angle here is related to the transverse momentum imbalance $q_\perp=|\vec p_{\rm T}^{\, jet}+\vec{p}_{\rm T}^{\, e}|$ in the plane transverse to the beam direction. The distribution peaks at zero as expected from LO DIS where the electron and jet are produced back-to-back. The finite width of the distribution is driven by the intrinsic $k_{\mathrm{T}}$ of the partons and gluon radiation. As shown by Liu et al.~\cite{Liu:2018trl}, in the limit that the transverse momentum imbalance $q_\perp$ is much smaller than the electron transverse momentum, this observable in e-p collisions provides clean access to the quark TMD PDF and to the Sivers effect when the proton is transversely polarized~\footnote{The Sivers effect refers to a correlation between the proton spin direction and the parton transverse momentum, which is quantified by the Sivers TMD PDF. It can be connected to parton orbital angular momentum.}. This measurement will be key for 3D imaging of the proton at the future EIC, which aims at understanding the nucleon in terms of quarks and gluons---a major goal of modern nuclear physics~\cite{Accardi:2012qut}. In particular this observable is insensitive to final state TMD effects, which provides a way to overcome the daunting task of simultaneously extracting TMD parton densities and fragmentation functions. We show the theoretical calculation of Ref.~\cite{Liu:2018trl} in Figure~\ref{fig:azimuthal} (solid black curve), which agrees well with the \textsc{Pythia8} simulation. 

A comparison of the cross section in e-p and e-A collisions is sensitive to \pTjet~broadening effects due to multiple scatterings in the medium. Such measurements are needed to quantify \qhat~in nuclei, as shown by Liu et al.~\cite{Liu:2018trl}. Following Refs.~\cite{Baier:1996sk,Liang:2008vz,Mueller:2016gko,Mueller:2016xoc}, the final state multiple scatterings of the struck quark/jet can be combined with the TMD distribution. Effectively, this leads to a modification of the resummed Sudakov exponent which can be expressed in terms of $\hat qL$. 

As we have shown in Section~\ref{sec:Simulations}, electron-jet correlations at the future EIC will sample $0.008<x<0.7$, which covers the shadowing, anti-shadowing and EMC regions. Electron-jet correlations in different kinematic bins will map these nuclear effects in 3D including potentially a parton flavor-separation. Azimuthal correlations provide a clean channel to explore nuclear tomography, extending traditional measurements based on hadrons~\cite{Qiu:2003pm,Dupre:2015jha}. 

A different definition of the transverse momentum measuring the imbalance between the electron and jet in semi-inclusive DIS was considered by Gutierrez-Reyes et al.~\cite{Gutierrez-Reyes:2019msa} This is sensitive to TMD PDFs and involves TMD evolution equations also for the final state jet. This observable can provide important complementary information for the nucleon and nuclear tomography.

\begin{figure*}
    \centering
    \includegraphics[width=1.0\textwidth]{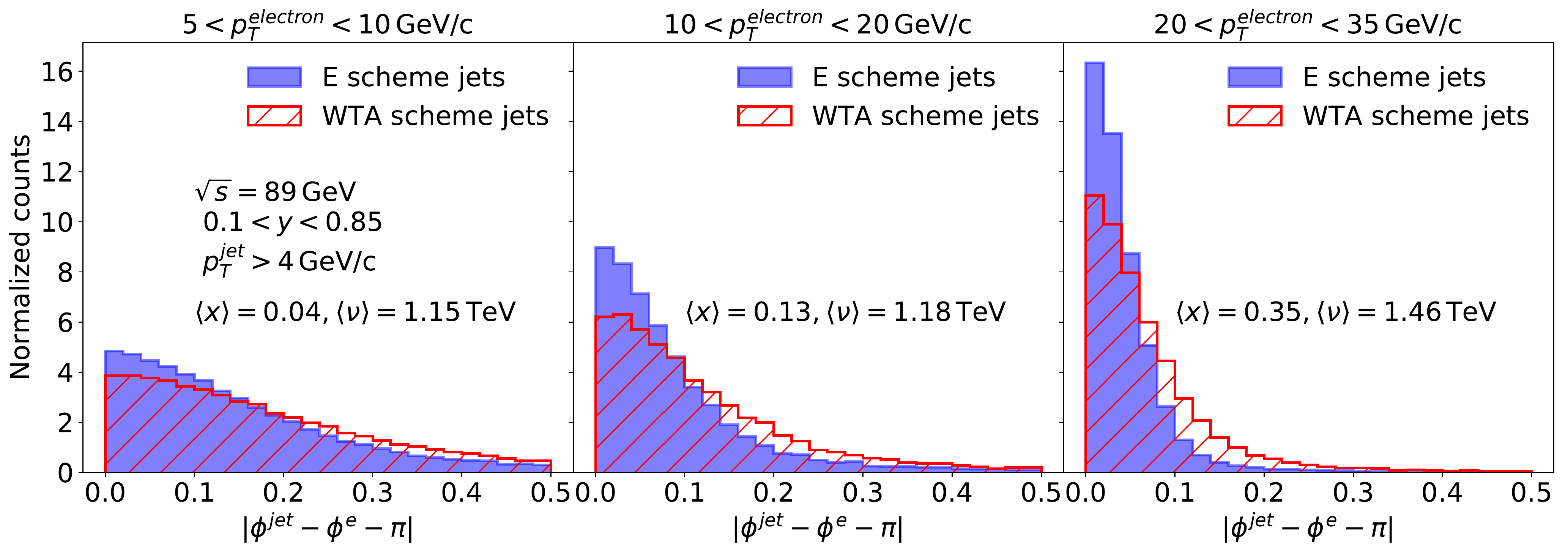}
\caption{Azimuthal angle correlation between the scattered electron and jets for the E-scheme and the winner-take-all (WTA) scheme. The different panels show different selections on \pTelectron. The projected statistical uncertainties are negligible and not shown.}
\label{fig:WTAjets}
\end{figure*}

The standard recombination scheme of jet reconstruction algorithms is the E-scheme, where at each step in the clustering the jet axis is defined by summing 4-vectors. The resulting jet axis is sensitive to recoil effects due to soft radiation in the jet. In contrast, the jet axis obtained with the winner-take-all scheme is by construction insensitive to soft radiation. At each step of the clustering, the jet axis is defined to be aligned with the more energetic particle. Therefore, this jet axis tracks collinear radiation.  

Recently, various observables involving the winner-take-all axis have been proposed~\cite{Neill:2016vbi,Gutierrez-Reyes:2018qez,Gutierrez-Reyes:2019vbx}. Potential applications include studies of the QGP, hadronization and studies of the intrinsic parton  $k_{\mathrm{T}}$ using jets in semi-inclusive DIS. In particular, comparisons between jets reconstructed with the standard E- and winner-take-all scheme in e-p and e-A collisions will shed light on the modification of collinear and/or soft fragmentation in nuclei and allow for quantitative studies of the jet broadening mechanism.

We consider the same observable as discussed in the previous section~\ref{sec:azimuthaldecorrelation} and investigate differences of the azimuthal angular correlation $|\phi^{jet} - \phi^e - \pi|$ between the electron and jet when the standard or winner-take-all jet axis is used. We note that as expected no significant difference between the \pTjet~spectra is observed since the clustering metric is the same for both recombination schemes. Figure~\ref{fig:WTAjets} shows the electron-jet azimuthal correlation for three intervals of \pTelectron~for E-scheme and winner-take-all jets. For both cases the distribution gets narrower with increasing \pTelectron. However, the winner-take-all jets show a significantly broader distribution for all \pTelectron~intervals. We expect these observables to be most relevant for studies of hadronization effects in e-p collisions and of broadening effects in e-A collisions. 
The significant difference between the standard and winner-take-tall axis observed here motivates further theoretical efforts in this direction.

\subsection{Groomed observables}
\label{sec:groomedjets}

Driven by LHC experiments, the field of jet substructure has grown rapidly in the last few years. See Ref.~\cite{Larkoski:2017jix,Asquith:2018igt} for recent reviews. An example is the 
shared momentum fraction, $z_{g}$, which is related to the Altarelli-Parisi splitting function~\cite{Larkoski:2017bvj} and is modified in heavy-ion collisions~\cite{Sirunyan:2017bsd,Acharya:2019djg}. The measurements rely on jet grooming algorithms such as ``soft drop''~\cite{Larkoski:2014wba}. Soft drop declustering isolates soft and wide-angle radiation inside the jet.
Nonperturbative effects such as hadronization corrections can be suppressed or enhanced depending on the observable under consideration, see for example recent work in Refs.~\cite{Makris:2017arq,Hoang:2019ceu,Chien:2019osu,Cal:2019gxa}. Furthermore, sensitivity to TMD PDFs can be improved~\cite{Gutierrez-Reyes:2019msa}.

The soft drop grooming algorithm operates on jets which are identified with the anti-k$_T$ algorithm and jet radius $R$. First, the jet is reclustered with the Cambridge/Aachen~\cite{Dokshitzer:1997in,Wobisch:1998wt} algorithm. This leads to an angular ordered clustering tree since, in this case, the pairwise distance metric only depends on the geometric distance between particles. Second, the jet is declustered recursively, and at each step the so-called soft drop condition is checked:
\begin{equation}
    \frac{{\rm min}[p_{T1},p_{T2}]}{p_{T1}+p_{T2}}>z_{\rm cut}\bigg(\frac{\Delta R_{12}}{R}\bigg)^\beta \,.
\end{equation}
Here, $p_{T1,2}$ denote the transverse momenta of the two branches of the jet at a given declustering step and $\Delta R_{12}$ is their geometric distance in the $\eta$-$\phi$ plane. $z_{\rm cut}$ and $\beta$ are free parameters that define the grooming procedure. If the branches fail this criterion, they are removed from the jet. Otherwise, the grooming algorithm terminates and returns the groomed jet which consists of the two branches that pass the criterion. The momentum sharing fraction $z_g$ and the groomed jet radius $R_g$ are defined in terms of the branches that pass the soft drop criterion:
\begin{equation}
    z_g=\frac{{\rm min}[p_{T1},p_{T2}]}{p_{T1}+p_{T2}}, \; R_g=\Delta R_{12} \,.
\end{equation}
The groomed radius $R_g$ corresponds to the opening angle between the two branches as the active area of the jet is given by $\sim\pi R_g^2$.

We anticipate that jet substructure and jet grooming will have an important role at the future EIC, just as at the LHC and RHIC, where it was used for 
tests of QCD in pp collisions and studies of the medium properties in p-A and AA collisions. For example, Ringer et al.~\cite{Ringer:2019rfk} showed that the groomed jet radius is sensitive to the jet transport coefficient similarly to electron-jet correlations. Probing the same physics with independent observables offers an important cross-check to ensure consistency and predictive power of theoretical calculations, and can be used in global extractions of \qhat. We expect that other observables will allow for similar studies where groomed jets can be used as well calibrated probes of nuclear effects in e-A collisions.

Here we study soft drop groomed jets at the future EIC focusing on the experimental feasibility of grooming low \pTjet~jets with modest constituent number. We use the \textsc{SoftDrop} algorithm~\cite{Larkoski:2014wba} as implemented in the \textsc{FastJet} package~\cite{Cacciari:2011ma}. The typical \pTjet~used in jet grooming studies at the LHC is ${\cal O}(100~\text{GeV/c})$~\cite{Aaboud:2017qwh,Sirunyan:2018xdh} but at the future EIC the range will be $\approx 10-35$~GeV/c, which is similar to the range explored at RHIC in p-p collisions~\cite{Adam:2020kug} ($15<\pTjet<60$ GeV/c for anti-$k_{\mathrm{T}}$ jets with $R=$0.4). The particle multiplicities in e-p collisions are smaller than in p-p. Consequently, we investigate how many particles are groomed away and how large the transverse momentum difference is before and after grooming at the future EIC. We choose the grooming parameters $z_{\mathrm{cut}}$=0.1, and  $\beta=0,\, 2$, which are often used in experimental studies at the LHC. Varying $\beta$ offers a way to explore different QCD dynamics and to gauge the sensitivity to soft radiation. The choice of $\beta$=0 ($\beta=2$) corresponds to more (less) aggressive grooming.

\begin{figure}
    \centering
    \includegraphics[width=1.0\columnwidth]{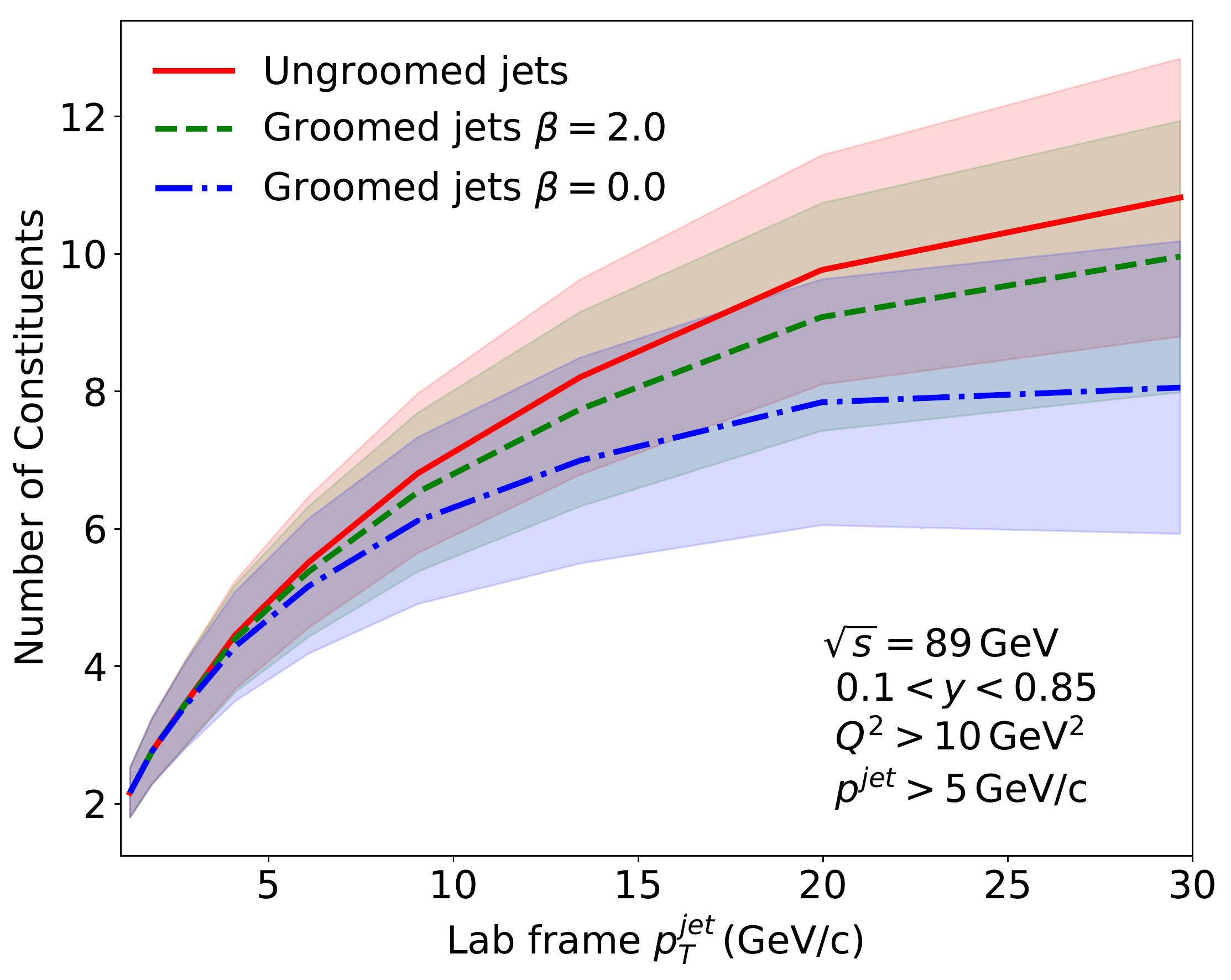}
\caption{Average number of jet constituents before and after grooming. The width represents the standard deviation of the distribution for each \pTjet~interval, where \pTjet referes to the transverse momentum of the ungroomed jet.}
\label{fig:groomed_particles}
\end{figure}

\begin{figure}
    \centering
    \includegraphics[width=1.0\columnwidth]{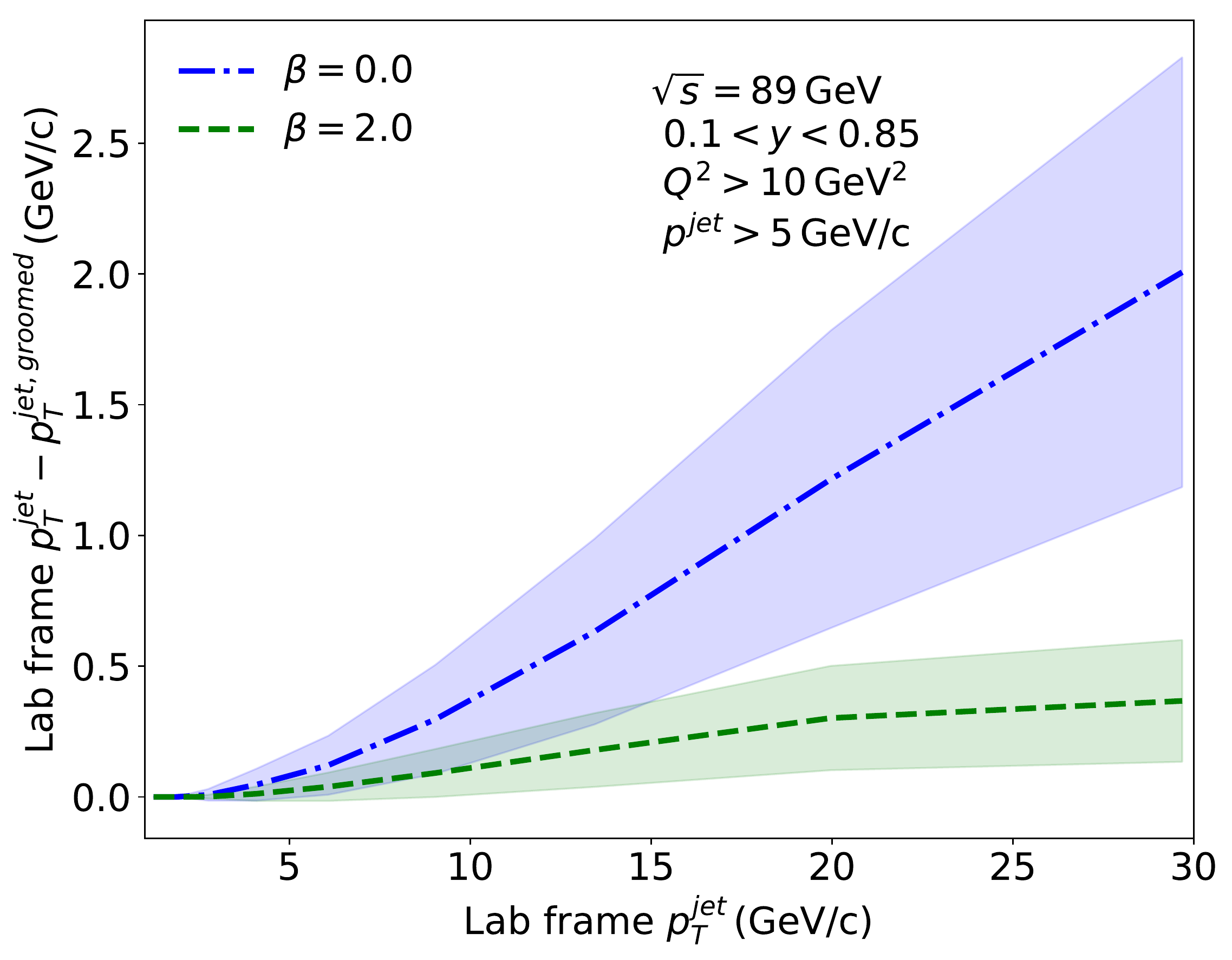}
\caption{Average \pTjet~removed by soft drop grooming with $\beta =0 $ and 2, as a function of the ungroomed \pTjet. The bands represent the standard deviation of the distribution. The jets are reconstructed with the anti-$k_{\mathrm{T}}$ algorithm with $R=1.0$.}
\label{fig:groomedpt}
\end{figure}

Figure~\ref{fig:groomed_particles} shows the number of particles in jets as a function of the ungroomed \pTjet~with and without grooming. The difference grows with \pTjet~and it reaches about $\approx 2$ particles on average for the $\beta=0$ case and $\approx 0.5$ particles for $\beta=2$. Figure~\ref{fig:groomedpt} shows the \pTjet~that is removed from the jet by the grooming procedure for the two grooming parameters $\beta=0,\,2$. We observe that the average value grows roughly linearly with ungroomed \pTjet~and at 30 GeV/c it reaches $\approx 2.0$ GeV/c for $\beta=0$ and $\approx 0.2$ GeV/c for $\beta=2$. We note that the standard deviation is large with respect to the average value, which indicates large fluctuations when using groomed jets. From Figures~\ref{fig:groomed_particles} and~\ref{fig:groomedpt} we conclude that the prospects of performing grooming at the future EIC, even with $\beta=0$, look promising. Depending on the observable under consideration it can be advantageous to choose a larger value of $z_{\rm cut}$ in order to extend the regime where perturbative calculations are applicable. Detailed detector simulation to quantify measurement effects on groomed variables is an important next step as well as detailed comparisons to theoretical calculations. 

\begin{figure*}
\centering
\includegraphics[width=0.66\textwidth]{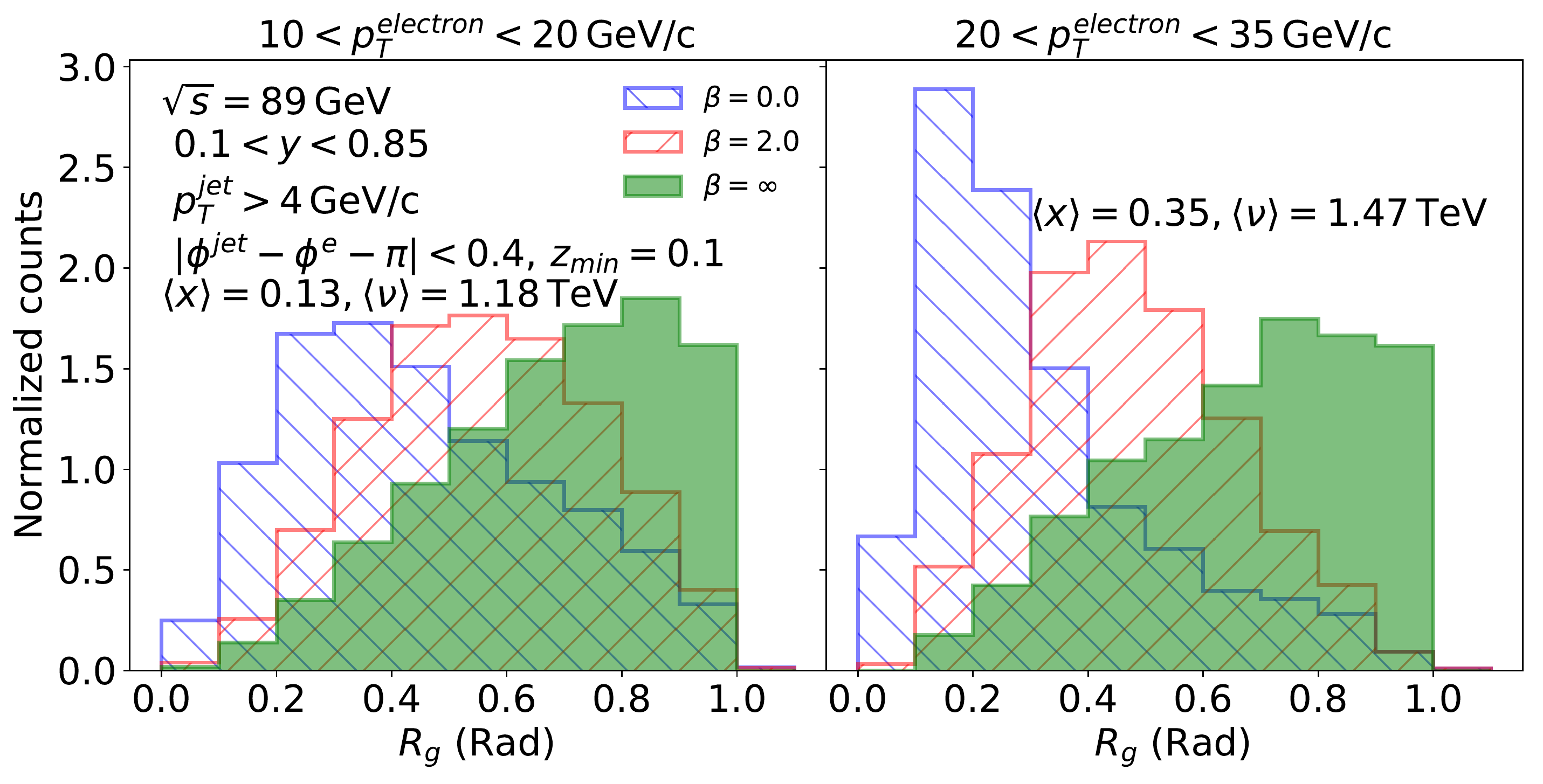}
\caption{Groomed jet radius for jets recoiling against the scattered electron for different \pTelectron~intervals and $\beta$ values. Here $\beta=0$ corresponds to the most aggressive grooming setup, whereas $\beta=\infty$ does not groom away any particle.}
\label{fig:softdrop_Rg}
\end{figure*}

Recent work by Ringer et al.~\cite{Ringer:2019rfk} showed that the jet groomed radius $R_g$, or equivalently, the angle between the two branches that pass the soft drop requirement, provides a new opportunity to investigate jet broadening effects. It is orthogonal to other observables that use more traditional jet variables such as the azimuthal angle and \pTjet. Figure~\ref{fig:softdrop_Rg} shows the groomed radius for jets recoiling against the scattered electron for two different \pTelectron~intervals. Here we consider the cases $\beta=0$ and $2$ as well as $\infty$. The limiting case of $\beta=\infty$ corresponds to no grooming and $R_g$ is the opening angle of the last two branches of the jet that were clustered together. The \Rg~distribution for $\beta=\infty$ is broad and peaks toward large values, with little dependence on \pTelectron. This distribution is dominated by power corrections and nonperturbative physics. Removing low momentum, wide-angle branches shifts the \Rg~distribution toward smaller values. As expected, $\beta=0$ yields a larger shift than $\beta=2$. We also observe that the shifts due to grooming are more significant for higher \pTelectron, which might be interpreted as a result of increased phase-space for soft radiation. 

In Ref.~\cite{Ringer:2019rfk}, it was proposed to study \pTjet~broadening effects in the QGP by considering the modification of the soft drop groomed jet radius. The same framework is applicable to studies of medium effects in the nucleus. In fact, the theory simplifies tremendously in e-p or e-A collisions because of the initial state electron and the large quark jet fraction. Here we work with the assumption of a pure quark jet sample; in the future this can be improved using the results of Ref.~\cite{Abelof:2016pby}. While the next-to-leading logarithmic corrections for this observable are known~\cite{Kang:2019prh}, we limit ourselves to a leading-logarithmic calculation~\cite{Larkoski:2014wba} as we are here mostly interested in the modification in e-A collisions. Nonperturbative hadronization effects are included through a convolution with a model shape function which depends on a single parameter. The size of hadronization corrections can be determined in e-p collisions by comparing to data or simulations, see~\cite{Ringer:2019rfk} for more details. 

\begin{figure}
    \centering
    \includegraphics[width=1.0\columnwidth]{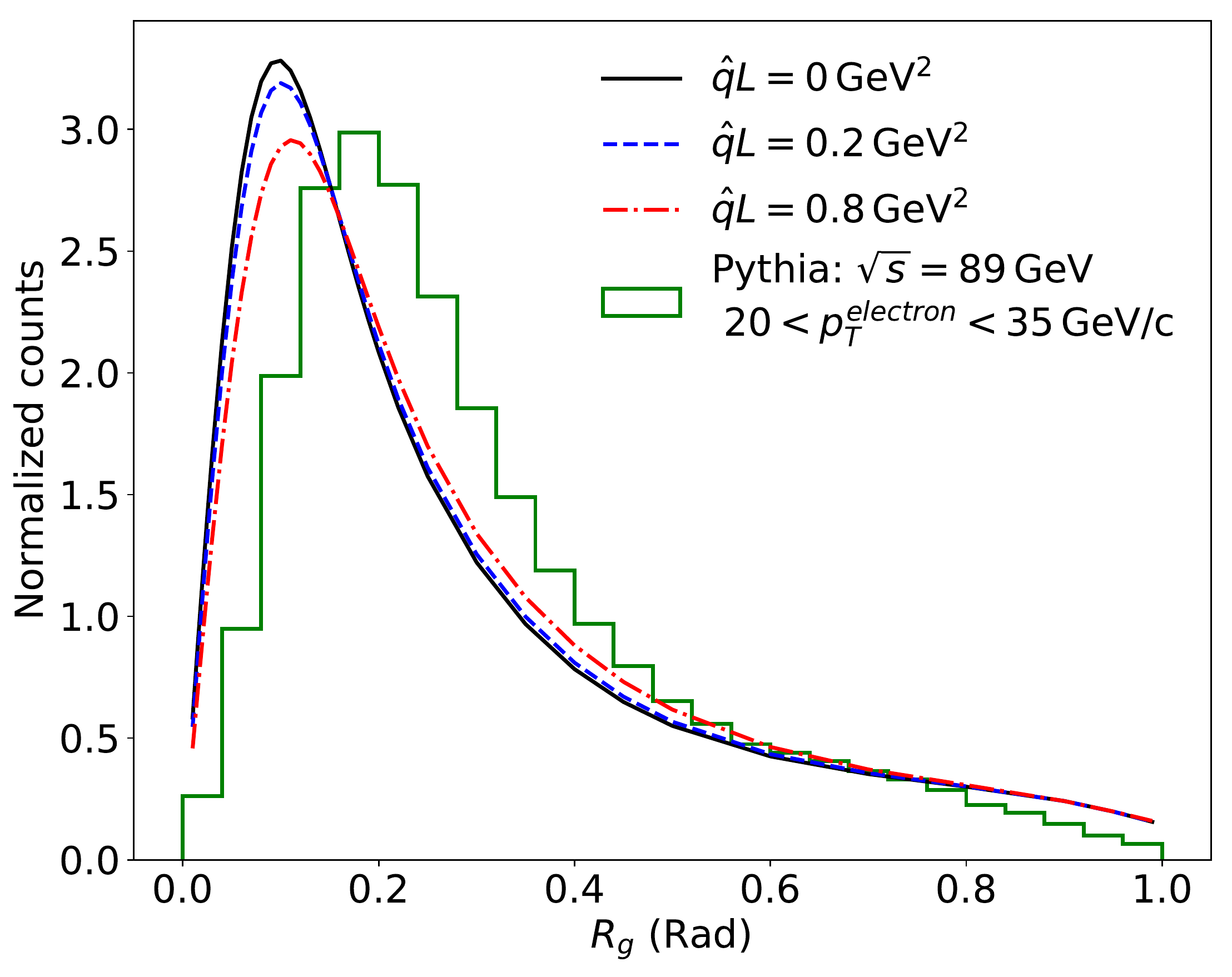}
    \caption{The soft drop groomed jet radius for e-p (solid black curve) and e-A collisions (dashed curves). The green histogram shows a \textsc{Pythia8} calculation for comparison. }
    \label{fig:qhat_groomed_radius}
\end{figure}

Figure~\ref{fig:qhat_groomed_radius} shows \textsc{Pythia8} results (green histogram) for $\beta=0$ and $20<\pTelectron<35$~GeV/c, which was also shown in the right panel of Figure~\ref{fig:softdrop_Rg}. The perturbative leading-logarithmic calculation of the groomed jet radius including hadronization effects (solid black curve) has a similar shape as the \textsc{Pythia8} results, 
though the \textsc{Pythia8} distribution is slightly shifted to the right. The other curves show the result when medium effects due to incoherent multiple scatterings of the two branches inside the nucleus are included. We parametrize the cold nuclear matter effects here analogously to the electron-jet azimuthal correlation~\cite{Liu:2018trl} considered in section~\ref{sec:azimuthaldecorrelation} above and choose the same values $\qhat L = 0.2$~GeV$^{2}$ and 0.8~GeV$^{2}$ (dashed curves) accordingly. The broadening effects are clearly visible and of similar magnitude as for the electron-jet azimuthal correlation observable. These results demonstrate that jet substructure observables offer novel and independent probes of nuclear effects at the future EIC.

\section{Experimental Aspects} \label{sec:experimentalaspects}

The modification of jet observables in e-A collisions compared to e-p are predicted to be at the few percent level. This places strict limits on systematic uncertainties of the measurements, and should inform detector designs for the future EIC. 

A disadvantage of jet measurements compared to single hadrons is that precise energy measurements are much more challenging. One of the most accurate jet energy measurements was performed by the ZEUS collaboration at HERA with its high resolution uranium-scintillator calorimeter,  yielding a jet energy scale (JES) uncertainty of $\pm1\%$ for jets with a transverse energy in the lab frame larger than 10 GeV~\cite{Abramowicz:2010ke}, and $\pm3\%$ for lower-energy jets. As jets have a rapidly falling spectrum, this energy scale uncertainty translates to an uncertainty of 5--10$\%$ for the \pTjet~spectra. Experiments at the LHC are close to achieving the goal of $\pm$1$\%$ JES as well. It seems unlikely that future EIC detectors will improve this. 

The JES uncertainty will thus likely be a limiting factor for jet measurements at the future EIC. Even for observables that do not require energy information per se, such as azimuthal differences between electrons and jets, the JES uncertainty enters as a second-order effect. For example, if a given observable depends on \pTjet, an unfolding procedure in more than one dimension will be needed. In particular, the azimuthal difference between jets and electrons has a rather strong \pTjet~dependence, as seen in Figure~\ref{fig:WTAjets}. 

Unlike fixed-target experiments that can use dual target techniques, data from e-A and e-p will be taken at different times and runs at a collider. Consequently, time-dependent changes in detector response will limit the cancellation in the e-A/e-p ratio and therefore drive the systematic uncertainties. 
Moreover, one of the most powerful calibration tools used by the HERA experiments was the momentum balance between the scattered electron and jets in neutral-current DIS~\cite{Newman:2013ada}. That effectively anchors the JES to the electromagnetic energy scale uncertainty, which is known much more precisely. That method is not available in for our tag and probe studies because it would use the same physics we want to study (at HERA, electron-jet correlations were primarily a calibration tool). This will increase the systematic uncertainty on the JES. 

Measuring ratios of cross sections in e-A and e-p collisions, allows some of the JES uncertainty to be canceled. In order to achieve an accuracy of 1$\%$ in \pTjet~spectra ratio measurements, one would need to reach a residual systematic uncertainty of 0.2$\%$ in the e-A/e-p ratio. We have shown that key observables such as the electron-jet azimuthal correlation and groomed-jet radius is rather sensitive to jet $\pTjet$ and that nuclear effects are predicted to be ${\cal O}(10\%)$ or less. Detailed detector simulations will be needed to see how residual JES uncertainties translate to systematic uncertainties in the e-A/e-p ratio for these observables; those studies should include realistic detector geometries, acceptance effects, as well as sensitivity to the modelling of the jet fragmentation pattern and its modification in e-A collisions. 

Geometrical acceptance considerations will also play an important role for the studies we suggest. For example, we have shown that the jets produced in the lowest $x$ events that can be reached will be produced predominantly in the region of $-2.0<\eta<-1.0$, as shown in Figure~\ref{fig:polarplot} b). This is a challenging region because it includes the transition between the barrel and endcap regions of a traditional collider experiment. Similarly, the high-\pT~jets that are produced in the highest-$x$ events cover the $1.0<\eta<2.0$ region, as shown in Figure~\ref{fig:polarplot} e). Further studies should address the degradation of performance due to material budget and any acceptance gaps to avoid missing this interesting jet kinematics. 

Another important point regards the potential need for a hadronic calorimeter. We have shown that low-$x$ events will produce low \pT~jets; those are better studied with tracks because calorimeters are limited by thresholds and the stochastic term of the energy resolution at low energies. Therefore, tracking efficiency and resolution will drive JES uncertainties. On the other hand, the measurement of high \pT~jets, which are produced mostly at mid rapidity  as shown in Figure~\ref{fig:polarplot} e), could benefit from a hadronic calorimeter. The impact of a mid-rapidity hadronic calorimeter has been recently studied by Page et al.~\cite{Page:2019gbf}, where it was shown that it could play an important role for accurate jet measurements as a neutral-hadron veto. Further studies on the interplay of tracking and calorimetry for jet measurements are needed to specify the requirements to measure the effects discussed in this paper.  

Jet substructure measurements also impose requirements on detector granularity. We have shown in Figure~\ref{fig:softdrop_Rg} that the groomed-jet radius at the future EIC peaks at $R_{g}\approx$~0.1 or larger, which is significantly larger than at the LHC. That is, the lower energy jets at future EIC are less collimated and thus impose less stringent requirements on detector granularity.  We foresee these measurements will be mainly based on the tracker and electromagnetic calorimeter, like the STAR measurements at RHIC~\cite{Adam:2020kug}. The expected granularity for both tracking and electromagnetic calorimeter of future EIC detectors~\cite{EICRandD} would likely not be an issue. Inclusion of hadronic calorimeter for jet constituents is possible, but may not be strictly necessary. For that, a granularity of at least $\Delta\eta \times \Delta\phi = 0.1\times0.1$ would be required. Further studies that translate realistic effects of detector granularity to resolution for $R_{g}$ and other jet substructure measurements would be informative.    

Finally, given that the future EIC jet measurements will likely be dominated by systematic uncertainties and the accuracy goal is at the percent-level, uncertainties due to luminosity and trigger efficiency will play a non-negligible role. We note that these are typically suppressed to the sub-percent level in fixed-target DIS experiments with the use of dual targets but in collider mode they will be non negligible. We again anticipate that the leading systematic uncertainty in the e-A/e-p ratios will be related to time-dependent effects in the trigger and luminosity calibrations. 

\section{Summary and Conclusions} \label{sec:conclusions}

We have explored the potential of jets at the future EIC as a precision tool for studies of the nucleus. We discussed requirements for semi-inclusive deep inelastic scattering ``tag and probe'' studies where the scattered electron fixes the jet kinematics, leading to an approach 
orthogonal to the HERA jet measurements, as well as to all previous projections of jet measurements at the future EIC. 

The kinematic reach for jet measurements at the future EIC is found to be roughly $0.008<x<0.7$ and $Q^{2}>25$ GeV$^{2}$ for $\sqrt{s}=89$ GeV. While the inclusive DIS measurements will have an extended kinematic reach, jets measurements will be indispensable for the study of quark-nucleus interactions, the quark-structure of nuclei in 3D, to tag the parton flavor and to separate current and target fragmentation.  

We identified several key observables for electron-jet studies, including the transverse momentum balance and the azimuthal angular correlation. We demonstrated the feasibility of groomed jets at the future EIC, to provide new tools for controlling hadronization effects. We presented comparisons to theoretical calculations where medium effects are included for both electron-jet correlations and jet substructure. Using information from different observables will be crucial to determine the jet transport coefficient $\hat q$. We also presented a study of the winner-take-all scheme for jet reconstruction, which will help to gauge the modification of soft and collinear fragmentation in the nucleus. 

Important future work includes studies with detector response simulations and more detailed comparisons to theoretical calculations. 

\section*{Acknowledgements} \label{sec:acknowledgements}
We thank Elke-Caroline Aschenauer and Brian Page for enlightening discussions about jet physics at the future EIC. We thank Frank Petriello for providing the NNLO DIS jet cross sections. We thank Alwina Liu, Stuti Raizada, and Feng Yuan for discussions and proofread on this manuscript. We thank Mateusz Ploskon for his technical help on interfacing \textsc{Pythia}, \textsc{FastJet} and \textsc{ROOT}. We also thank Jose Soria for his contributions in the early stage of this work. This work is supported by the US Department of Energy Office of Nuclear Physics, by the US National Science Foundation, and by the University of California, Office of the President. M.A acknowledges partial support through DOE Contract No. DE-AC05-06OR23177 under
which JSA operates the Thomas Jefferson National Accelerator Facility.


\renewcommand\refname{Bibliography}
\bibliography{biblio.bib} 

\appendix*

\end{document}

%% file: preamble.tex
\usepackage{amsthm}
\usepackage{mathtools}
\usepackage{physics}
\usepackage{xcolor}
\usepackage{graphicx}
\usepackage[left=23mm,right=13mm,top=35mm,columnsep=15pt]{geometry} 
\usepackage{adjustbox}
\usepackage{placeins}
\usepackage[T1]{fontenc}
\usepackage{lipsum}
\usepackage{csquotes}